%% file: neurips_2025.tex
\documentclass{article}

 \usepackage[dandb, final]{neurips_2025}

\usepackage[utf8]{inputenc} 
\usepackage[T1]{fontenc}    
\usepackage{hyperref}       
\usepackage{url}            
\usepackage{booktabs}       
\usepackage{amsfonts}       
\usepackage{nicefrac}       
\usepackage{microtype}      
\usepackage{xcolor}         
\usepackage{enumitem}
\usepackage{graphicx}
\usepackage{multirow}
\usepackage{pifont}   
\usepackage{amsmath}
\title{MRSAudio: A Large-Scale Multimodal Recorded Spatial Audio Dataset with Refined Annotations}

\author{
Wenxiang Guo$^{\dagger 1,2}$, Changhao Pan$^{\dagger 1}$, Zhiyuan Zhu$^{\dagger 1}$, Xintong Hu$^{\dagger 1}$, Yu Zhang$^{\dagger 1}$\and 
Li Tang$^{1}$, Rui Yang$^{1}$, Han Wang$^{1}$, Zongbao Zhang$^{1}$, Yuhan Wang$^{1}$, Yixuan Chen$^{1}$, Hankun Xu$^{1}$\and 
Ke Xu$^{1}$, Pengfei Fan$^{1}$, Zhetao Chen$^{1}$, Yanhao Yu$^{1}$, Qiange Huang$^{1}$, Fei Wu$^{1}$, Zhou Zhao$^{\ddagger 1,2}$\\[2pt]
$^{1}$Zhejiang University\quad
$^{2}$ Shanghai AI Laboratory\\
\texttt{\{guowx314,panch,zhaozhou\}@zju.edu.cn}\\[2pt]
\small $^{\dagger}$Equal contribution.\quad $^{\ddagger}$Corresponding author.
}

\begin{document}

\maketitle

\begin{figure}[h]
\centering
\includegraphics[width=\linewidth]{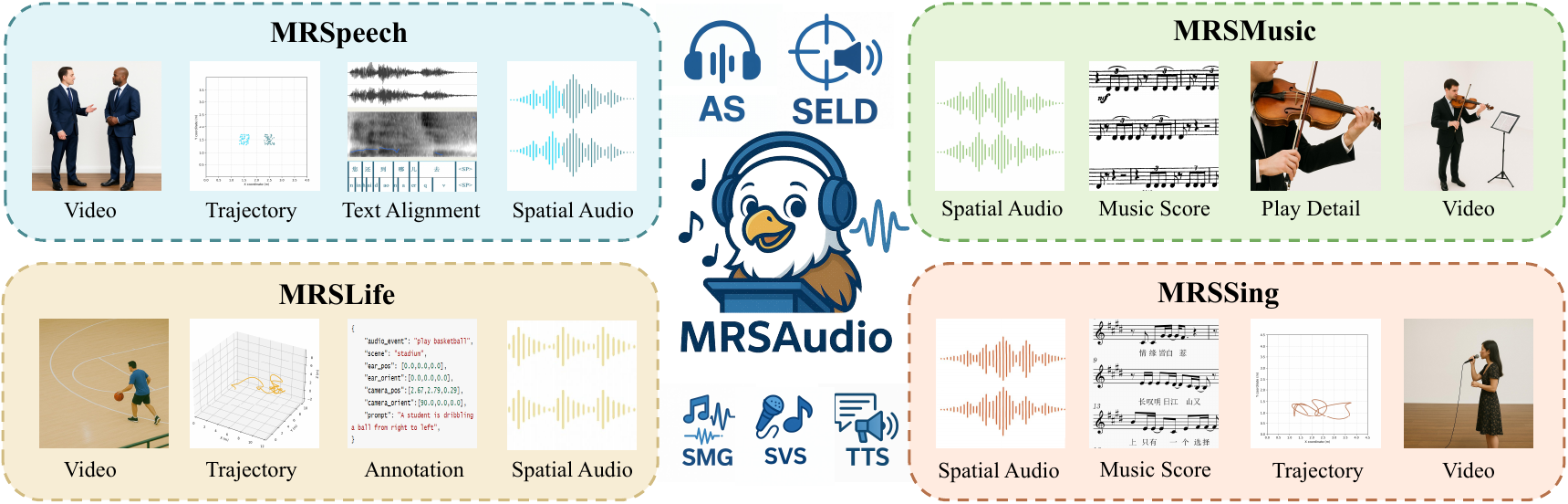}
\caption{
Overview of MRSAudio. The dataset comprises four real-world scenarios: MRSSpeech, MRSLife, MRSMusic, and MRSSing, each with multimodal annotations for spatial audio research.
}
\label{fig:com}
\end{figure}

\begin{abstract}


Humans rely on multisensory integration to perceive spatial environments, where auditory cues enable sound source localization in three-dimensional space. 
Despite the critical role of spatial audio in immersive technologies such as VR/AR, most existing multimodal datasets provide only monaural audio, which limits the development of spatial audio generation and understanding. 
To address these challenges, we introduce MRSAudio, a large-scale multimodal spatial audio dataset designed to advance research in spatial audio understanding and generation. 
MRSAudio spans four distinct components: MRSLife, MRSSpeech, MRSMusic, and MRSSing, covering diverse real-world scenarios. 
The dataset includes synchronized binaural and ambisonic audio, exocentric and egocentric video, motion trajectories, and fine-grained annotations such as transcripts, phoneme boundaries, lyrics, scores, and prompts.
To demonstrate the utility and versatility of MRSAudio, we establish five foundational tasks: audio spatialization, and spatial text to speech, spatial singing voice synthesis, spatial music generation and sound event localization and detection. 
Results show that MRSAudio enables high-quality spatial modeling and supports a broad range of spatial audio research.
Demos and dataset access are available at \url{https://mrsaudio.github.io}.
\end{abstract}

\input{Sections/1_introduction}

\input{Sections/2_relatedwork}

\input{Sections/3_dataset}
\input{Sections/4_benchmark}
\input{Sections/5_conclusion}

\begin{ack}
This work was supported by the National Key R\&D Program of China (2022ZD0162000), the National Natural Science Foundation of China under Grant No. 62222211, and the National Natural Science Foundation of China under Grant No. U24A20326.

\end{ack}

\newpage
\bibliographystyle{neurips_2025}
\bibliography{neurips_2025}


\newpage
\appendix
\input{Sections/6_appendix}




\end{document}

%% file: Sections/1_introduction.tex
\section{Introduction}
\label{sec: intro}

Humans rely on multisensory integration to perceive and interpret physical environments. With the rapid growth of film, virtual reality (VR), augmented reality (AR), and gaming applications, users increasingly expect not only precise audiovisual alignment but also highly immersive experiences. While recent advances in deep learning have enabled realistic generation of speech, music, and sound effects synchronized with text or video \citep{kreuk2022audiogen, yang2023uniaudio, wang2024frieren}, most models focus on monaural audio and neglect spatialized soundscapes that enhance immersion. The human binaural system uses interaural time differences (ITD) and interaural level differences (ILD) to localize sound in three-dimensional space, and spatial audio must remain consistent with visual cues \citep{yost1998spatial, cohen1999maps, grothe2010mechanisms}. Any mismatch can disrupt immersion and weaken the sense of presence. For example, hearing a cat’s meow from the left immediately suggests its location, even if it is just off-screen.

Despite the growing importance of spatial audio in these immersive technologies\citep{xie2020spatial, huiyu2025psa}, progress in machine learning for spatial audio understanding is limited by the fundamental spatial data constraints. Most existing audio datasets\citep{jort_audioset_2017, chen2020vggsound, agostinelli2023musiclm} focus on monaural recordings, which discard vital spatial information, effectively "flattening" the soundscape and preventing models from learning key physical phenomena such as room reverberation, echo patterns, and sound propagation. Moreover, the scarcity of multimodal datasets that align spatial audio with synchronized visual, position geometric, and semantic annotations hinders the development of advanced auditory scene-analysis systems capable of human-like spatial perception.

To address these gaps, we present \textbf{MRSAudio}, a 484-hour large-scale multimodal spatial audio dataset designed to support both spatial audio understanding and generation. It integrates high-fidelity spatial recordings with synchronized video, 3D pose tracking, and rich semantic annotations, enabling comprehensive modeling of real-world auditory scenes. As shown in Figure~\ref{fig:com}, the dataset comprises four subsets, each targeting distinct tasks and scenarios.
\textbf{MRSLife} (129 h) captures daily activities such as board games, cooking, and office work, using egocentric video and FOA audio annotated with sound events and speech transcripts.
\textbf{MRSSpeech} (206 h) includes binaural conversations from 44 speakers across diverse indoor environments, paired with video, 3D source positions, and complete scripts.
\textbf{MRSSing} (80 h) features high-quality solo singing performances in Chinese, English, German, and French by 20 vocalists, each aligned with time-stamped lyrics and corresponding musical scores.
\textbf{MRSMusic} (69 h) offers spatial recordings of 23 Traditional Chinese, Western and Electronic instruments, with symbolic score annotations that support learning-based methods for symbolic-to-audio generation and fine-grained localization.
Together, these four subsets support a broad spectrum of spatial audio research problems, including event detection, sound localization, and binaural or ambisonic audio generation. By pairing spatial audio with synchronized exocentric and egocentric video, geometric tracking, and detailed semantic labels, MRSAudio enables new research directions in multimodal spatial understanding and cross-modal generation. Throughout this paper, unless stated otherwise, we use the term spatial audio to refer to binaural audio.

\begin{itemize}[leftmargin=*]
  \item We introduce \textbf{MRSAudio}, a 484-hour, large-scale multimodal spatial audio dataset explicitly designed to push the boundaries of spatial audio understanding and generative modeling.
  \item We assemble synchronized binaural and ambisonic recordings with exocentric and egocentric video, geometric source positions, transcripts, scores, lyrics, and event labels, providing one of the most richly annotated multimodal resources for spatial audio research.
  \item We organize the data into four complementary subsets (MRSLife, MRSSpeech, MRSSing, MRSMusic), each carefully tailored to different real-world acoustic scenarios and equipped with rich, scenario-specific annotations to facilitate downstream task development.
  \item We establish and release evaluation protocols and baseline implementations for five benchmark tasks: audio spatialization generation, spatial text-to-speech, spatial singing voice synthesis, spatial music generation, and localization and detection of sound events, in order to demonstrate MRSAudio’s versatility and to foster reproducible research.
\end{itemize}

The remainder of this paper is organized as follows.  
Section~\ref{sec: rel} reviews existing spatial audio datasets.  
Section~\ref{sec: data} describes the design, collection process, and key statistics of MRSAudio.  
Section~\ref{sec: ben} presents extensive benchmark experiments using state-of-the-art methods on five core spatial audio tasks: audio spatialization, spatial text-to-speech, spatial singing voice generation, spatial music synthesis, and sound event localization and detection.  
Finally, Section~\ref{sec: con} concludes the paper and discusses the limitations and potential risks associated with MRSAudio.

%% file: Sections/2_relatedwork.tex
\section{Related Work}
\label{sec: rel}
Deep learning has driven remarkable progress in both audio generation \citep{huang2023makeanaudio,yang2023uniaudio,kreuk2022audiogen} and audio understanding \citep{Qwen-Audio,huang2023audiogpt,tang2023salmonn} tasks. However, the majority of these advances still rely heavily on monaural audio, which lacks the ability to represent or capture the rich spatial cues that naturally occur in real-world environments.
The rapid adoption of VR/AR technologies has concurrently driven growing demand for immersive spatial audio experiences. Researchers have focused on several key technologies including: sound event localization and detection \citep{Adavanne2019_DCASE,wang2022nerc}, mono-to-spatial audio conversion \citep{Gao20192.5D,morgadoNIPS18}, and end-to-end spatial audio generation \citep{sun2024both,kimvisage,liu2025omniaudiogeneratingspatialaudio}. 

Despite recent progress, spatial audio generation and understanding remain constrained by the paucity of high-quality datasets. Due to the difficulty and expense of collecting and annotating high-quality spatial audio datasets, most open-source datasets still primarily consist of simulated or web-crawled content.
Simulated datasets offer precise annotations but lack perceptual realism, while crawled datasets often provide real-world diversity but are missing critical labels, such as listener and source positions and content-level annotations, thus limiting their utility.
For instance, spatial symbolic music generation demands accurate musical scores and corresponding positional metadata.
To better understand the current landscape, we survey existing spatial audio datasets. These datasets differ in terms of audio format, including First-Order Ambisonics (FOA), multi-channel arrays, and binaural microphones, as well as in their collection methods (simulated, crawled, recorded) and annotation. 

\begin{table}[ht]
\centering
\small
\caption{Comparison of spatial audio datasets, where T denotes speech transcripts, P represents sound source positions, N indicates natural language prompts, and C stands for sound class tags}
\scalebox{0.9}{
\begin{tabular}{l|ccc|c|c|c|c|cccc}
\toprule
\multirow{2}{*}{\bfseries Dataset} & \multicolumn{3}{c|}{\textbf{Audio Format}} & \multirow{2}{*}{\textbf{Collect}} & \multirow{2}{*}{\textbf{Hours}} & \multirow{2}{*}{\textbf{Type}} & \multirow{2}{*}{\textbf{Visual}} & \multicolumn{4}{c}{\textbf{Label}} \\
 & FOA & Multi & Binaural & & & & & T & P & N & C \\
\midrule
Spatial LibriSpeech & \ding{51} & \ding{51} & \ding{55} & Simulated & 650 & Speech & -     & \ding{51} & \ding{51} & \ding{55} & \ding{55} \\
YT-Ambigen          & \ding{51} & \ding{55} & \ding{55} & Crawled   & 142 & ALL  & Video & \ding{55} & \ding{55} & \ding{55} & \ding{55} \\
BEWO-1M             & \ding{55} & \ding{55} & \ding{51} & Craw+Sim  & 2800 & ALL  & Image & \ding{55} & \ding{55} & \ding{51} & \ding{55} \\
FAIR-Play           & \ding{55} & \ding{55} & \ding{51} & Recorded  & 5.2 & Music  & Video & \ding{55} & \ding{55} & \ding{55} & \ding{55} \\
STARSS23            & \ding{51} & \ding{51} & \ding{55} & Recorded  & 7.5 & ALL  & Video & \ding{55} & \ding{51} & \ding{55} & \ding{51} \\
BinauralMusic       & \ding{55} & \ding{55} & \ding{51} & Crawled   & 15.2 & Music & Video & \ding{55} & \ding{55} & \ding{55} & \ding{51} \\
RealMAN       & \ding{55} & \ding{51} & \ding{55} & Recorded   & 228.2 & Speech & Image & \ding{51} & \ding{51} & \ding{55} & \ding{55} \\
Sphere360           & \ding{51} & \ding{55} & \ding{55} & Crawled   & 288 & ALL  & Video & \ding{55} & \ding{55} & \ding{55} & \ding{55} \\
\textbf{MRSAudio (Ours)} & \ding{51} & \ding{55} & \ding{51} & Recorded & 484 & ALL & Video & \ding{51} & \ding{51} & \ding{51} & \ding{51} \\
\bottomrule
\end{tabular}}
\label{tab:data}
\end{table}

As shown in Table~\ref{tab:data}, existing datasets vary in their goals and modalities. For instance, Spatial LibriSpeech \citep{sarabia2023spatial} simulates spatial speech using the LibriSpeech corpus and is mainly intended for binaural TTS applications. RealMAN \citep{yang2024realman} is a large-scale real-recorded and annotated 32-channel microphone array dataset for multichannel speech enhancement and source localization. YT-Ambigen \citep{kimvisage} and Sphere360 \citep{liu2025omniaudiogeneratingspatialaudio} are derived from web-crawled video datasets, but lack explicit spatial annotations, making them suitable primarily for video-to-spatial audio generation. BinauralMusic focuses on musical content, offering instrument class tags. BEWO-1M \citep{sun2024both} combines crawled content and simulated binaural rendering, and provides image or GPT-generated prompts. STARSS23 \citep{shimada2023starss23} features real-world FOA and multi-channel audio alongside synchronized videos and includes sound class and position labels.
In contrast to prior datasets, MRSAudio offers a comprehensive, large-scale, and real-world spatial audio corpus, featuring 484 hours of recorded data in both FOA and binaural formats, covering all audio domains including general audio, speech, singing, and music. Uniquely, MRSAudio includes synchronized audio, video, and 3D positional geometry, along with fine-grained cross-modal annotations, such as: transcripts, word and phoneme boundaries and music scores.
These features make MRSAudio an ideal resource for a broad range of spatial audio generation and understanding tasks, including spatial speech, music, and singing voice synthesis.

%% file: Sections/3_dataset.tex
\section{Dataset Description}
\label{sec: data}

\begin{figure}[ht]
\centering
\includegraphics[width=\linewidth, trim={0mm 30mm 0mm 40mm}, clip]{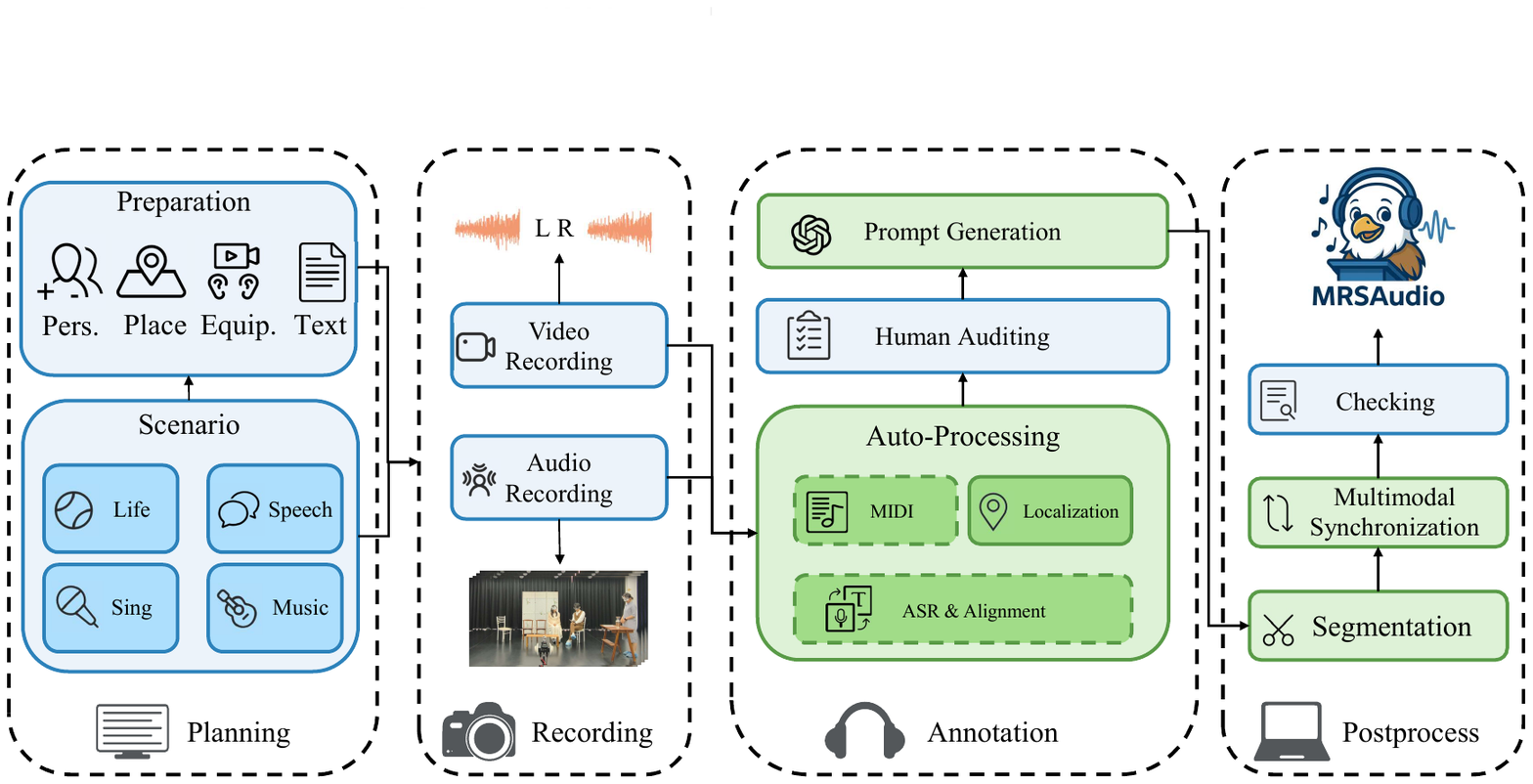}
\caption{
The pipeline of data collection and processing of MRSAudio.
The blue boxes indicate steps requiring manual intervention, while the green boxes denote automated processing. In the “auto-processing” section, dashed modules apply to some scenarios, while solid modules apply to all.
}
\label{fig: pipeline}
\end{figure}

In this section, we present \textbf{MRSAudio}, a freely available multimodal spatial audio corpus with synchronized video, positional data, and fine-grained annotations, released under the CC BY-NC-SA 4.0 license. Figure \ref{fig: pipeline} illustrates our data processing pipeline, with detailed descriptions in subsequent subsections. We then summarize key statistics that demonstrate MRSAudio's scale and diversity.

\subsection{Planning}
To ensure that MRSAudio comprehensively covers scenarios from daily life, speech, singing, and music, we conduct systematic and modular planning before recording as follows. 

\textbf{MRSLife:} This subset focuses on everyday conversations and environmental sound events. Based on the degree of human vocal interaction, MRSLife is further divided into two parts: MRSDialogue, which captures unscripted conversations that naturally include spontaneous action sounds (e.g., footsteps, door movements), and MRSSound, which focuses on non-verbal sound events primarily caused by physical activities such as cooking, typing, or sports.

\textbf{MRSSpeech:} MRSSpeech targets clean, high-quality conversational recordings for TTS task. All spoken interactions are recorded in controlled indoor environments with minimal noise. We invite speakers to participate in content-driven conversations based on predefined scripts. 

\textbf{MRSSing:} MRSSing captures solo vocal performances for singing voice synthesis tasks. It includes professional solo vocal recordings in four languages: Mandarin, English, German, and French, performed by singers covering the full vocal range including soprano, alto, tenor, and bass.

\textbf{MRSMusic:} MRSMusic captures immersive instrumental performances suitable for spatial music generation and analysis.
We record solo performances from 45 professional musicians across 23 instruments, including Traditional Chinese, Western and Electronic instruments. 
Each performance is paired with its corresponding musical score to support score-based music generation.

Details regarding personnel recruitment, venue selection, equipment configuration, and material preparation for each subset are provided in Appendix~\ref{app: pl}.

\subsection{Recording}
Based on the predefined planning, we conduct parallel data collection across all modules.
To ensure participant anonymity, masks are worn during recording when necessary. All participants sign an open-source data release agreement, allowing the dataset to be freely distributed for academic research purposes.
The recording details are summarized as follows:

\textbf{MRSLife:} In MRSDialogue, audio is recorded using a professional binaural recording head and high-resolution sound cards, while synchronized video is captured using industry-standard cameras. In MRSSound, in addition to binaural audio and exocentric video, we also captured FOA (Zoom H3-VR) and egocentric video (Gopro camera).To ensure the effectiveness of the egocentric  video, participants are asked to remain within the frontal field of view of the binaural recording head.

\textbf{MRSSpeech:} To introduce spatial variability, we select four recording rooms that differ in size and acoustic material. Each speaker reads inflected emotions from the scripts while walking through the space, producing dynamic spatial cues. Recordings include spoken passages, binaural audio and exocentric video, as well as a clean mono audio by a lavalier microphone placed near the speaker.

\textbf{MRSSing:} The professional singers perform according to musical scores. To introduce variation in source–listener geometry, we adjust the position of the head-mounted binaural microphone relative to the singer. In addition to the binaural audio and synchronized video, each session includes a clean vocal track recorded with a studio-grade condenser microphone.

\textbf{MRSMusic:} We record 23 Traditional Chinese, Western and Electronic instruments, performed by 45 
professional musicians.To capture rich spatial detail, we vary microphone placement around the instrument. Recordings include binaural audio, monaural audio, exocentric video, and synchronized video that records playing gestures. Full recording details are provided in Appendix \ref{app: re}.

\subsection{Annotation}
To maximize MRSAudio’s utility across a wide range of tasks, we begin with event-level annotations for all vocal and acoustic content in MRSLife, MRSSpeech, MRSSing, and MRSMusic. However, coarse annotations alone are insufficient for fine-grained tasks such as singing voice modeling and music generation from scores. To bridge this gap, we design a comprehensive annotation pipeline. Full implementation details and detailed annotation guidelines are provided in Appendix \ref{app: an}.

\textbf{MRSLife:}
For MRSDialogue, we apply WhisperX \citep{bain2023whisperx} for automatic speech recognition and speaker diarization to generate initial transcripts and speaker turns. Human annotators correct recognition errors and speaker attribution mismatches. The audio is then segmented into utterances and the transcripts are converted into phoneme sequences (using pypinyin for Mandarin). A two-stage alignment process follows: we first apply the Montreal Forced Aligner (MFA) \citep{mcauliffe2017montreal} for coarse word/phoneme mapping, then manually refine boundaries in Praat \citep{boersma2001praat}.  
For MRSSound, we annotate sound event categories and corresponding time intervals.

\textbf{MRSSpeech:}
Given the availability of full scripts, we adapt WhisperX for long-form word-to-audio alignment of up to 30 minutes.
Each script line is automatically matched to its corresponding audio segment (see Appendix \ref{app: an} for more details). 
Annotators then review these alignments, correcting any omissions or insertions caused by actors' deviations from the script. Finally, phoneme sequences are extracted and aligned with the audio using the same procedure as in MRSDialogue.

\textbf{MRSSing:}
We use voice activity detection (VAD) to segment recordings into singing regions, then align pre-existing lyrics using LyricFA’s ASR-based dynamic programming algorithm\footnote{\url{https://github.com/wolfgitpr/LyricFA}}.
Phoneme generation is language-dependent: pypinyin for Mandarin, ARPA for English, and MFA’s built-in phoneme sets for French and German. Alignment is conducted via MFA, followed by manual refinement. Melody and rhythm of singing are transcribed into MIDI format using ROSVOT \citep{li2024robust}. Annotators then label each excerpt with high-level style descriptors, such as emotional tone (happy, sad), tempo (slow, moderate, fast), and pitch range (low, medium, high).

\textbf{MRSMusic:}
We use Audio Slicer\footnote{\url{https://github.com/flutydeer/audio-slicer}} to segment the music recordings and generate initial symbolic annotations with basic-pitch \citep{2022_BittnerBRME_LightweightNoteTranscription_ICASSP}, and then employ professional musicians to verify and adjust note onsets, offsets, and dynamics to match the performance accurately.

\textbf{Source Localization:}
For static sources, we manually record 3D positions relative to the capture space. For dynamic scenes, we use the Ultra-Wideband (UWB) system\citep{aiello2003ultra} to track the positions of sound sources in real time. Based on the recorded position trajectories, we generate natural language motion descriptions using GPT-4o \citep{achiam2023gpt}.

\subsection{Post-Processing}
The post-processing pipeline comprises three key steps to refine raw annotated data. First, segmentation splits continuous recordings into task-oriented clips: utterances for speech and singing are extracted via alignment timestamps, while MRSSound audio is uniformly divided into 10-second segments. Next, multimodal synchronization aligns each clip with auxiliary modalities (text, video, position metadata) with temporal anchors. Static sound sources are annotated with manually measured 3D coordinates, whereas dynamic sources leverage interpolated UWB tracking trajectories. For scenes where participants’ faces are visible, anonymization is performed by adding half-face masks. Finally, quality assurance involves domain experts auditing 15\% of clips across all modules to verify temporal alignment precision, cross-modal content consistency, and annotation accuracy.
Full auditing protocols are documented in Appendix~\ref{app: post}, ensuring reproducibility of this process.

\subsection{Statistics}

\begin{figure}[ht]
\centering
\includegraphics[width=\linewidth, trim={0mm 40mm 0mm 0mm}, clip]{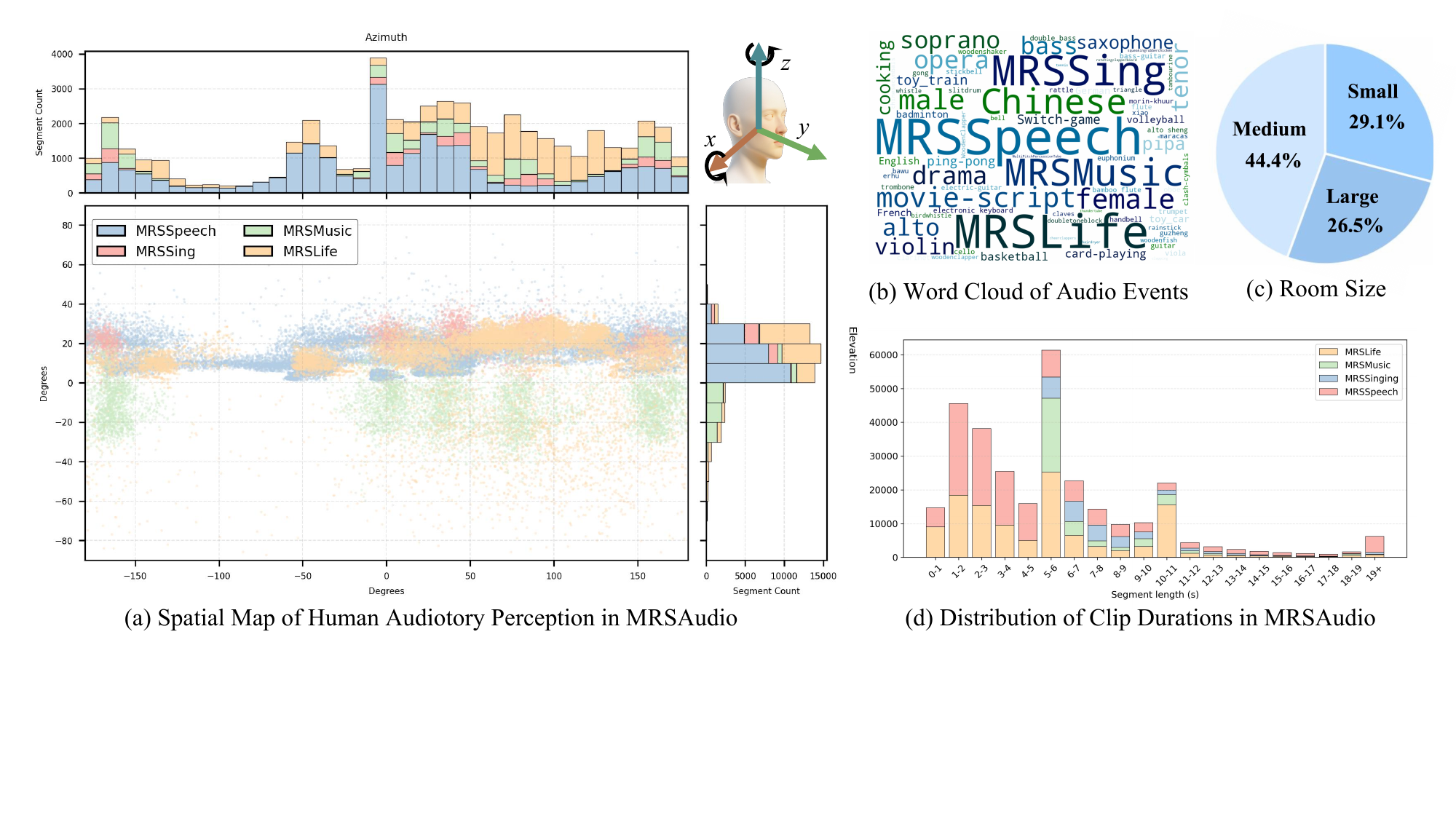}
\caption{
Statistical overview of MRSAudio. (a) Spatial distribution of sound sources relative to the listener. Red, green, and blue arrows denote the positive x, y, and z axes; azimuth is measured around the z-axis from the x-axis, and elevation is relative to the xy-plane. (b) Word cloud. (c) Proportions of recording spaces by room size. (d) Distribution of audio segment durations. 
}
\label{fig:stat}
\end{figure}

To illustrate spatial diversity, Figure~\ref{fig:stat}(a) shows the 3D distribution of sound source positions relative to the listener. The heatmap reveals near-uniform azimuthal coverage, with greater density in the frontal hemisphere due to the prevalence of egocentric video recordings. Elevation angles are concentrated between –40° and 40°, aligning with typical human sound perception patterns.
Figure~\ref{fig:stat}(b) summarizes the annotations using a keyword cloud that captures the range of real-world activities recorded. Figure~\ref{fig:stat}(c) presents the proportions of recording spaces by room size: medium-sized rooms are the most common, accounting for approximately 40\% of sessions, while small and large rooms each represent around 30\%. Recording duration is evenly distributed across room types, ensuring diverse acoustic coverage.
Figure~\ref{fig:stat}(d) shows the distribution of segment durations after automatic and manual segmentation. Most audio clips are shorter than 10 seconds, which is suitable for modeling short-duration events and supports efficient downstream training and inference.
Overall, MRSAudio offers comprehensive coverage across spatial positions, acoustic environments, scene types, and temporal structures, making it well-suited for a wide range of spatial audio generation and understanding tasks. A more detailed breakdown of per-scenario statistics is provided in Appendix~\ref{app: stat}.

%% file: Sections/4_benchmark.tex
\section{Benchmarks}
\label{sec: ben}

To demonstrate the quality and utility of MRSAudio in real-world scenarios, we evaluate it on five representative spatial audio tasks: (i) audio spatialization, (ii) spatial text-to-speech, (iii) spatial singing voice synthesis, (iv) spatial music generation, and (v) sound event localization and detection (SELD). These tasks cover both generation and understanding, and are critical for applications such as AR/VR, spatial media production, and perceptual scene analysis.
All experiments are conducted using state-of-the-art methods on a server equipped with eight NVIDIA RTX 4090 GPUs. 
For different tasks, we employ distinct objective metrics. For generation tasks, we compute cosine similarity scores for direction (ANG Cos) and distance (DIS Cos) to quantify spatial alignment quality. Additionally, we utilize subjective MOS-Q (Mean Opinion Score for Quality) to evaluate the quality of generated audio and MOS-P (Mean Opinion Score for Position) to assess spatial perception. For implementation details of  training and evaluation metrics, please refer to Appendix~\ref{app: exp}.

\subsection{Audio Spatialization}
\label{sec:bag}

Audio spatialization aims to synthesize spatially immersive soundscapes from monaural inputs and source positional information. To evaluate MRSAudio on this task, we adopt BinauralGrad~\citep{leng2022binauralgrad}, a diffusion-based generation model that predicts binaural waveforms conditioned on monaural audio and source coordinates. Since the downstream generation tasks can be formulated as predicting interaural (binaural) representations from mono audio, we employ a single-stage training scheme for all experiments. For comparison, we include a traditional signal processing baseline (DSP), which renders binaural audio using virtual source positions simulated via room impulse responses (RIRs) and head-related transfer functions (HRTFs).
We adopt the following objective metrics to evaluate audio quality:  
(1) W-L2: waveform L2 distance,  
(2) A-L2: amplitude envelope L2 distance,  
(3) P-L2: phase difference L2,  
(4) STFT: multi-resolution STFT loss,  
(5) PESQ: perceptual speech quality score.
Results for the ground truth and DSP baseline are obtained by averaging across all test sets from the four MRSAudio subsets. Further details are provided in Appendix~\ref{app:bas}.

\begin{table}[ht]
\centering
\small
\caption{Audio Spatialization Performance. For MRSlife, we only use the MRSSound subset.}
\label{tab:bag}
\begin{tabular}{l|ccccc|cc}
\toprule
\bfseries Method   & \textbf{W-L2 $\times 10^{-3}$} $\downarrow$ & \textbf{A-L2} $\downarrow$ & \textbf{P-L2} $\downarrow$ & \textbf{PESQ} $\uparrow$ & \textbf{STFT} $\downarrow$ & \textbf{MOS-Q} $\uparrow$ & \textbf{MOS-P} $\uparrow$ \\
\midrule
Ground Truth       & –                 & –             & –               & –             & –              &  4.69 $\pm$ 0.08      &  4.56 $\pm$ 0.10 \\
\midrule
DSP                & 1.691              & 0.048          & 1.562            & \bf 2.830          & \bf 1.246           & 3.89 $\pm$ 0.09      & 3.75 $\pm$ 0.11 \\
MRSLife  & 0.076             & 0.025         &  0.898            & -          & 2.243           & 3.91 $\pm$ 0.07      &3.87 $\pm$ 0.10 \\
MRSSpeech & 0.460         & 0.061          & 0.807            & 1.929          & 2.352          & 3.88  $\pm$  0.08 & 3.84  $\pm$  0.08 \\
MRSSing    & 0.647         & 0.093          & 1.004            & 1.723          & 2.539  & 3.84 $\pm$ 0.09 & 3.91 $\pm$ 0.07 \\
MRSMusic   & 0.705         & 0.063          & 0.835            & -          & 1.724   & 3.87 $\pm$ 0.07 & 3.93 $\pm$ 0.09 \\
ALL    & \bf 0.305         & \bf 0.041          & \bf 0.801           & 2.352          & 1.681  & \bf 3.93 $\pm$ 0.09 & \bf    3.94 $\pm$ 0.07 \\
\bottomrule
\end{tabular}
\end{table}

As shown in Table~\ref{tab:bag}, BinauralGrad achieves strong performance across all MRSAudio subsets, surpassing the classical DSP baseline on most objective and subjective metrics. The highest performance is observed on MRSLife, likely due to the relatively limited variation in sound sources within these scenes. These results demonstrate that MRSAudio’s rich spatial annotations and diverse acoustic environments provide a solid foundation for training and evaluating spatial audio generation models.

\subsection{Spatial Text to Speech}
\label{sec: btts}
Spatial TTS aims to produce high-quality speech enriched with spatial cues, thereby enhancing immersion and realism in AR/VR. Although recent advances in TTS have led to impressive improvements in speech quality, progress in spatialized speech generation remains limited due to the scarcity of spatially annotated recordings with rich, high-quality labels.
To evaluate the effectiveness of MRSAudio for this task, we train an end-to-end model following  ISDrama \citep{zhang2025isdrama} to directly generate spatial speech from text and position data. Additionally, we compare with cascaded pipelines that combine monaural TTS models (CosyVoice~\citep{du2024cosyvoice} and F5-TTS~\citep{chen2024f5}) with the audio spatialization module. This allows us to assess both speech generation quality and the spatial fidelity enabled by our dataset.
For content evaluation, we report Character Error Rate (CER) and Speaker Similarity (SIM). Further details are provided in Appendix~\ref{app:btts}.

\begin{table}[ht]
\centering
\small
\caption{Spatial TTS Performance on MRSSpeech. “SP” denotes the Audio Spatialization.}
\scalebox{1}{
\begin{tabular}{l|cccc|cc}
\toprule
\multirow{2}{*}{\bfseries{Method}} & \multicolumn{4}{c|}{\bfseries Objective} & \multicolumn{2}{c}{\bfseries Subjective} \\
 & CER $\downarrow$ & SIM $\uparrow$ & ANG Cos $\uparrow$ & DIS Cos $\uparrow$ & MOS-Q $\uparrow$ & MOS-P $\uparrow$ \\
\midrule
Ground Truth & \bf 2.54\% & –     & –    & –    & 4.39 $\pm$ 0.08      & 4.16 $\pm$ 0.10 \\
\midrule
Mono + SP         & 2.56\% & \bf 0.98 & 0.44 & \bf 0.68 & \bf 3.88  $\pm$  0.08 & \bf 3.84  $\pm$  0.08 \\
CosyVoice + SP    & 3.89\% & 0.96 & 0.41 & 0.63 & 3.75  $\pm$  0.12 & 3.72  $\pm$  0.09 \\
F5-TTS + SP       & 3.15\% & 0.97 & 0.40 & 0.62 & 3.69  $\pm$  0.13 & 3.67  $\pm$  0.14 \\
ISDrama (speech)    & 3.35\% & 0.96 & \bf 0.48 & 0.65 & 3.85  $\pm$  0.09 & 3.82  $\pm$  0.11 \\
\bottomrule
\end{tabular}}
\label{tab:tts}
\end{table}

As shown in Table~\ref{tab:tts}, the Mono+SP method achieves strong performance across most metrics. The CER remains low and comparable to the ground truth, indicating preserved linguistic accuracy after spatialization. A high SIM score reflects stable timbre learning. ANG Cos and DIS Cos show good spatial alignment with the ground truth, and subjective MOS scores confirm that the generated speech is both natural and spatially coherent. These results demonstrate that MRSSpeech provides high-quality, spatially annotated training data that enables effective and realistic spatial TTS.

\subsection{Spatial Singing Voice Synthesis}
\label{sec: bsvs}

Spatial SVS aims to produce expressive, high-quality singing voices enriched with accurate spatial cues, thereby enhancing listener immersion. While traditional SVS has advanced considerably, spatial SVS remains underexplored due to the lack of high-quality, spatially annotated datasets.
To evaluate MRSAudio for this task, we use the MRSSing subset to train and benchmark models. We adopt the ISDrama architecture to perform end-to-end spatial singing synthesis, incorporating note-level pitch control to enhance prosody accuracy. For comparison, we use the open-source SVS models Rmssinger~\citep{he2023rmssinger}. The outputs are then spatialized with BinauralGrad.
We employ objective metrics, including Mel-Cepstral Distortion (MCD) and F0 Frame Error (FFE), to evaluate spectral and pitch similarity between predicted and ground-truth. Details are in Appendix~\ref{app:bsvs}.
\begin{table}[ht]
\centering
\small
\caption{Spatial SVS Performance on MRSSing. “SP” denotes the Audio Spatialization.}
\scalebox{1}{
\begin{tabular}{l|cccc|cc}
\toprule
\multirow{2}{*}{\bfseries{Method}} & \multicolumn{4}{c|}{\bfseries{Objective}} & \multicolumn{2}{c}{\bfseries{Subjective}}\\
 & MCD $\downarrow$ & FFE $\downarrow$ & ANG Cos $\uparrow$ & DIS Cos $\uparrow$ & {MOS-Q $\uparrow$} & {MOS-P $\uparrow$} \\
\midrule
GT (Ground Truth) & - & - & - & - & 4.45 $\pm$ 0.10 & 4.30 $\pm$ 0.12 \\
\midrule
Mono+SP &  \bf 3.19 & \bf 0.17 & \bf 0.51 & \bf 0.71  & 3.84 $\pm$ 0.09 & \bf 
 3.91 $\pm$ 0.07 \\
Rmssinger+SP & 3.85 & 0.23 & 0.45 & 0.65 & 3.65 $\pm$ 0.08 & 3.81 $\pm$ 0.11 \\
ISDrama(sing) & 3.71 & 0.21 & 0.47 & 0.70 & \bf 3.86 $\pm$ 0.13 & 3.88 $\pm$ 0.09 \\
\bottomrule
\end{tabular}}
\label{tab: bsvs}
\end{table}

As shown in Table~\ref{tab: bsvs}, the Mono + SP approach trained on MRSSing with pitch control achieves the best performance across most metrics. Its low MCD indicates strong spectral fidelity, and high ANG Cos and DIS Cos scores demonstrate effective spatial alignment. Subjective MOS results confirm that the generated singing voices are natural, high-quality, and spatially coherent.
These findings validate MRSSing as an effective resource for spatial singing voice generation.

\subsection{Spatial Music Generation}
\label{sec: bmg}

This task aims to synthesize spatially immersive music conditioned on symbolic scores. While datasets like FAIR-Play offer high-quality instrument recordings, they lack aligned sheet music, limiting controllable generation. In contrast, our MRSMusic subset includes 23 instruments with aligned scores, enabling fine-grained, score-based spatial music synthesis.
We benchmark three systems on MRSMusic. First, we apply BinauralGrad~\citep{leng2022binauralgrad} to spatialize mono recordings. Second, we utilize Make-An-Audio 2~\citep{copet2023simple} to generate mono music from MIDI scores, which is subsequently spatialized. Third, we adapt ISDrama to accept both score embeddings and spatial poses, enabling end-to-end spatial symbolic music generation.
For objective evaluation, we compute Fréchet Audio Distance (FAD) and FFE to evaluate the results. Details are in Appendix~\ref{app:smg}.

\begin{table}[ht]
\centering
\small
\caption{Spatial MG Performance on MRSMusic. “SP” denotes the Audio Spatialization.}
\label{tab:bmg}
\begin{tabular}{l|cccc|cc}
\toprule
\multirow{2}{*}{\bfseries Method} & \multicolumn{4}{c|}{\bfseries Objective} & \multicolumn{2}{c}{\bfseries Subjective} \\
 & FAD $\downarrow$ & FFE $\downarrow$ & ANG Cos $\uparrow$ & DIS Cos $\uparrow$ & MOS-Q $\uparrow$ & MOS-P $\uparrow$ \\
\midrule
GT (Ground Truth)          &    –   &    –   &    –   &    –   &4.49 $\pm$ 0.09 & 4.34 $\pm$ 0.11 \\
\midrule
Mono+SP         & 2.88   & \bf 0.14   & 0.48   & \bf 0.74    & 3.87 $\pm$ 0.07 & \bf 3.93 $\pm$ 0.09 \\
Make-An-Audio 2+SP & 4.39   & 0.49   & 0.49   & 0.41  & 3.74 $\pm$ 0.10 & 3.73 $\pm$ 0.13 \\
ISDrama(music)  & \bf 2.45   & 0.21   &  \bf 0.53   &  0.68   &  \bf 3.89 $\pm$ 0.12 & 3.88 $\pm$ 0.10 \\
\bottomrule
\end{tabular}
\end{table}
As shown in Table~\ref{tab:bmg}, the Mono + SP pipeline achieves strong performance, benefiting from access to ground truth mono audio. The Make-An-Audio 2 + SP exhibits limitations in FAD and pitch accuracy. This suggests that general-purpose audio generation models may struggle to fully capture structured musical information from symbolic prompts. The ISDrama (Music) model generates music directly from the symbolic inputs and spatial cues, achieving better coherence and spatial alignment than the Make-An-Audio 2 + SP. These results highlight MRSMusic’s effectiveness in supporting spatially controllable music generation across diverse instruments and spatial conditions.

\subsection{Sound Event Localization and Detection}
\label{sec: seld}
This task evaluates the ability to detect and localize sound events using MRSAudio’s spatial annotations. We follow STARSS23~\citep{shimada2023starss23} using both audio-only and audio-visual variants on the MRSSound subset.
In the audio-only condition, models receive either FOA or binaural waveforms as input and predict sound event classes along with 3D source coordinates. For the audio-visual condition, we extract bounding boxes of visible persons to serve as coarse visual priors, which are fused with the audio representations. We also explore architectural variations by replacing the original convolutional backbone with a Transformer encoder, allowing us to assess the impact of temporal modeling capacity on spatial prediction.
We evaluate model performance using four standard joint detection and localization metrics, including location-aware detection ($F_{20^\circ}$, $ER_{20^\circ}$) and class-aware localization ($LE_{CD}$, $LR_{CD}$). Details are in Appendix~\ref{app:seld}.

\begin{table}[ht]
\centering
\small
\caption{Sound Event Localization and Detection on MRSSound.}
\label{tab:seld}
\begin{tabular}{l|c|c|cccc}
\toprule
\bfseries Model & \bfseries Audio Type & \bfseries Visual & ER$_{20^\circ}$ $\downarrow$ & F$_{20^\circ}$ $\uparrow$ & LE$_{CD}$ $\downarrow$ & LR$_{CD}$ $\uparrow$ \\
\midrule
ConvNet     & FOA       & \ding{51} & 1.17 $\pm$ 0.02 & \bf 13.00 $\pm$ 2.72 & 42.44 $\pm$ 8.91  & 82.76 $\pm$ 5.19 \\
ConvNet     & FOA       & \ding{55} & 1.12 $\pm$ 0.02 & 9.33 $\pm$ 0.36  & 46.47 $\pm$ 4.46  & 85.35 $\pm$ 3.80 \\
ConvNet     & Binaural  & \ding{51} & 1.11 $\pm$ 0.03 & 5.90 $\pm$ 3.86  & 41.95 $\pm$ 9.85  & 45.81 $\pm$ 11.29 \\
ConvNet     & Binaural  & \ding{55} & 1.11 $\pm$ 0.02 & 6.00 $\pm$ 0.34  & 45.17 $\pm$ 7.92  & 70.14 $\pm$ 10.38 \\
Transformer & FOA       & \ding{51} & 1.01 $\pm$ 0.01 & 7.52 $\pm$ 0.42  & \bf 35.69 $\pm$ 7.47  & 46.78 $\pm$ 7.62 \\
Transformer & FOA       & \ding{55} & \bf 0.99 $\pm$ 0.05 & 7.95 $\pm$ 0.15  & 48.76 $\pm$ 2.66  & \bf 87.32 $\pm$ 2.14 \\
Transformer & Binaural  & \ding{51} & 1.01 $\pm$ 0.02 & 8.47 $\pm$ 0.65  & 36.18 $\pm$ 7.29  & 41.33 $\pm$ 12.73 \\
Transformer & Binaural  & \ding{55} & 1.11 $\pm$ 0.04 & 7.37 $\pm$ 0.97  & 46.74 $\pm$ 7.91  & 43.87 $\pm$ 10.74 \\
\bottomrule
\end{tabular}
\end{table}

As shown in Table~\ref{tab:seld}, model performance varies with architecture, input modality, and audio format. Transformer-based models generally outperform ConvNet baselines, particularly in reducing localization error. FOA input consistently yields better results than binaural audio, benefiting from richer spatial representation. For example, the Transformer with FOA and no visual input achieves the lowest error rate (ER$_{20^\circ}$ 0.99) and highest localization recall (LR$_{CD}$ 87.32). The addition of visual features improves performance in some ConvNet settings (e.g., F$_{20^\circ}$ rises from 9.33 to 13.00 with FOA), but offers limited or inconsistent gains in Transformer models, possibly due to modality mismatch or redundancy. These results highlight the value of MRSAudio’s spatial annotations and multimodal streams in supporting flexible evaluation under both unimodal and multimodal configurations.

%% file: Sections/5_conclusion.tex
\section{Conclusion and Discussion}
\label{sec: con}
We introduce MRSAudio, a large-scale, multimodal spatial audio corpus designed to support a wide range of generation and understanding tasks. MRSAudio comprises four complementary modules, MRSLife, MRSSpeech , MRSSing, and MRSMusic, each captured with binaural/FOA audio, synchronized video, precise 3D pose metadata, and richly detailed annotations (event labels, transcripts, phoneme boundaries, lyrics, musical scores, and motion prompts). Through extensive benchmarks on audio spatialization, binaural speech, singing, and music generation , as well as sound event localization and detection, we demonstrate that MRSAudio’s scale, diversity, and annotation depth enable state‐of‐the‐art performance and unlock new avenues for spatial audio research.

\textbf{Limitations and Future Directions:}  
While MRSAudio offers broad multimodal coverage, two limitations remain. First, although synchronized video is provided for all recordings, current benchmarks only explore visual input in a limited subset of tasks. Second, while the dataset includes both binaural and First-Order Ambisonic (FOA) formats, the FOA subset is smaller in scale, and most tasks focus on binaural audio, limiting spatial modeling diversity.
Future work will expand the role of visual modalities in tasks such as sound localization and scene understanding. We also plan to increase FOA recordings to balance data distribution and support broader spatial audio research, and develop more FOA-specific benchmarks to better utilize ambisonic spatial cues.

\textbf{Negative Societal Impact:}  
As with any large-scale audiovisual dataset, MRSAudio carries potential risks if misused. It could be exploited to generate highly realistic yet synthetic spatial audio for deepfakes or disinformation in AR and VR applications. To mitigate such risks, MRSAudio is released under a noncommercial license with clear usage guidelines. We encourage responsible use in accordance with ethical standards, including consent management, data governance, and transparency.

%% file: Sections/6_appendix.tex
\section{Details of Dataset}
\label{app: stat}
\subsection{Details of Planning}
\label{app: pl}
During the planning phase, we systematically analyze real-world application scenarios for spatial audio and divide them into four major categories. MRSLife focuses on daily environmental sound events, MRSSpeech targets high-quality conversational data, MRSSing captures solo vocal performances, and MRSMusic records immersive instrumental music. These four scenarios collectively cover a broad range of everyday acoustic contexts and are designed to support a wide spectrum of downstream tasks.

Based on predefined scenario requirements, we recruit professionals from relevant fields to participate in the recording process. To ensure spatial diversity, we rent various venues tailored to each scenario type. Following the FAIR-Play protocol, we use a 3Dio Free Space XLR binaural microphone to capture spatial audio, GoPro HERO cameras and mobile phones to record exocentric and egocentric videos, and UWB-based tracking systems to capture motion trajectories.
Once personnel, locations, and equipment are secured, we prepare task-specific materials for each subset. For MRSLife, we further divide the content based on the proportion of speech involved, resulting in two subcategories: MRSDialogue (e.g., group games, board games) and MRSSound (e.g., kungfu, kitchens, offices, sports). We predefine common sound events for each, such as clattering dishes, keyboard typing, and table tennis.
For MRSSpeech, we compile a large corpus of scripts from movies, TV shows, and crosstalk performances and automatically extract dialogue passages for speaker delivery.
For MRSSing, we design lyrics in four languages (Chinese, English, German, and French) and recruit singers across a range of vocal types to maximize diversity.
For MRSMusic, we collect solo performances across 23 traditional and modern instruments, covering a wide array of timbres and playing techniques.
Specific sound categories for each subset are summarized in Table~\ref{tab:categories}.

\begin{table}[ht]
\centering
\small
\caption{Examples of predefined content types for each MRSAudio subset.}
\label{tab:categories}
\begin{tabular}{l|p{11cm}}
\toprule
\textbf{Subset} & \textbf{Samples of Categories or Keywords} \\
\midrule
\textbf{MRSLife} & \textbf{MRSDialogue}: board games, card games, collaborative tasks
\newline
\textbf{MRSSound}: kungfu, office, maracas, typing, whistle, gong \\
\midrule
\textbf{MRSSpeech} & Movie scripts, crosstalk, scripted TV dialogue, multi-speaker conversations \\
\midrule
\textbf{MRSSing} & Chinese, English, German, French; soprano, alto, tenor, bass \\
\midrule
\textbf{MRSMusic} & violin, electronic keyboard, cello, viola, double bass, trumpet, trombone, euphonium, erhu, pipa, xiao, bawu, trumpet, trombone, and others (23 instruments in total) \\
\bottomrule
\end{tabular}
\end{table}

\subsection{Details of Recording}
\label{app: re}
We recruit a large number of participants with professional backgrounds in singing, music, and language to contribute to the recording process. To protect their identity, participants are asked to wear masks in appropriate scenarios. Prior to participation, all individuals sign consent forms agreeing to the open-source release of their audio and video under the CC BY-NC-SA 4.0 license.

All audio is recorded in WAV format at a sampling rate of 48 kHz. Video is recorded at a minimum resolution of 1080p and 24 frames per second, and is later standardized to this format during post-processing.

\textbf{MRSLife:}
We recruit 62 participants to perform daily activities including board games, cooking, exercise, and office work. In MRSDialogue, each participant is compensated \$30 per recorded hour. Binaural audio is captured using head-mounted microphones, and third-person video is recorded. In MRSSound scenes such as kung fu or kitchen demonstrations, performers are paid \$50 per hour. Both binaural and FOA (First-Order Ambisonics) recordings are collected, along with first-person and third-person video. The binaural dummy head is co-located with the egocentric camera and the Zoom H3-VR (Ambisonic recorder). The egocentric camera is rigidly aligned with the head's gaze to capture first-person visuals, while the Ambisonic recorder is mounted 7.5 cm above the head. An exocentric camera is placed at a surveyed position in the scene to provide a third-person view of the environment and object relationships.
Participants receive brief scene descriptions and are asked to act naturally while maintaining a single active sound source when possible. The total duration for MRSLife recordings reaches approximately 150 hours.

\textbf{MRSSpeech:}
We employ 44 expressive speakers to read from scripted texts, each paid \$30 per hour of recorded audio. The total recorded duration reaches 200 hours, with compensation totaling \$40,000. During recording sessions, speakers alternate between standing and walking around the recording area to introduce spatial diversity while maintaining speech clarity. Binaural audio is captured using head-mounted microphones, and clean monaural speech is recorded using lavalier microphones. All sessions are filmed from a third-person perspective.

\textbf{MRSSing:}
Eighteen professional singers participate in the recordings. Each singer is fluent in at least one of the following languages: Chinese, English, German, or French, and collectively they cover all vocal ranges including soprano, alto, tenor, and bass. Performers are paid \$50 per hour, contributing a total of 80 hours of audio. Singing is recorded at a fixed position to ensure spatial consistency. Binaural recordings are captured with head-mounted microphones, and monaural audio is recorded with a studio-grade microphone. Third-person panoramic video is also captured.

\textbf{MRSMusic:}
We engage 45 instrumentalists performing 23 Traditional Chinese, Western and Electronic instruments such as violin, erhu, pipa, electric guitar, and keyboard. Each performer receives \$60 per recorded hour. A total of 69 hours of solo music performances are recorded. Audio is captured using both head-mounted binaural microphones and reference monaural microphones. Third-person video provides a full-scene view, while first-person cameras are used to capture detailed playing gestures.

\subsection{Details of Annotation}
\label{app: an}
We employ a team of domain experts in singing, music performance, and linguistics to carry out and review all annotations, compensating each annotator at a rate of \$15 per hour. Prior to beginning their work, every expert receives a clear explanation of how the annotations will be used and agrees to release their annotation results under an open‐source license for academic research.
For all modules, we first synchronize each audio with its corresponding video, mono‐channel reference track, and 3D positional metadata. Annotators then verify and correct this synchronization to ensure perfect alignment across modalities.

\textbf{MRSLife.}  
In MRSDialogue scenes, we automatically generate initial transcripts and speaker clusters with WhisperX, extracting word‐level timestamps and speaker IDs. Experts then load these results in Praat and assign each cluster to the correct speaker, correcting transcription errors as needed. Next, we run Montreal Forced Aligner (MFA) \citep{mcauliffe2017montreal} to produce coarse phoneme‐to‐audio alignments (exported in TextGrid format), using pypinyin to convert Chinese text into phoneme sequences.\footnote{\url{https://github.com/mozillazg/python-pinyin}} Finally, annotators refine word and phoneme boundaries in Praat to achieve millisecond‐level precision. 
In MRSSound segments, annotators additionally label each time interval with the corresponding event category (e.g., “clattering,” “typing,” “pages turning”).

\textbf{MRSSpeech.}  
Given full dialogue scripts, we perform automatic alignment to 30-minute recordings using a chunk-based extension of WhisperX. This method divides each utterance-level audio segment into fixed-length chunks (e.g., 30 seconds), applies phoneme-level models (e.g., wav2vec 2.0 \citep{baevski2020wav2vec} ) for emission prediction on each chunk, and then concatenates emissions to form a complete alignment matrix.
We compute alignments via a trellis-based dynamic programming algorithm with backtracking, followed by scaling to restore absolute timestamps. Word- and sentence-level timings are derived by grouping aligned characters using word indices and sentence tokenization. Missing or partial timings are interpolated using a specified method (e.g., nearest).
This pipeline enables efficient and accurate alignment of long-form recordings on GPU. We then refine sentence boundaries in Praat and apply the same MFA-plus-Praat workflow used in MRSLife for fine-grained phoneme and word-level alignment.

\textbf{MRSSing.}  
Each segment contains a solo vocal performance with known lyrics. We first apply voice activity detection (VAD) to segment the recordings. Lyrics are aligned to each segment using LyricFN’s ASR-based dynamic programming method. English phonemes follow the ARPABET standard,\footnote{\url{https://en.wikipedia.org/wiki/ARPABET}} while German and French use MFA’s phoneme sets.\footnote{\url{https://mfa-models.readthedocs.io/en/latest/dictionary/}} We then apply MFA for initial alignment and refine it manually in Praat to obtain precise word and phoneme boundaries. Annotators assign fine-grained style tags, including emotion, genre, and tempo. To generate score annotations, we extract F0 contours using RMVPE~\citep{wei2023rmvpe} and convert them into MIDI format via ROSVOT~\citep{li2024robust}, followed by expert correction.

\textbf{MRSMusic.}  
We segment the recordings using AutoSlicer~\citep{liu2022autoslicer} and generate preliminary symbolic annotations using basic-pitch~\citep{2022_BittnerBRME_LightweightNoteTranscription_ICASSP}. Professional musicians then verify and refine the annotations, adjusting note onsets, offsets, dynamics, and articulations to ensure consistency between the score and performance.

\subsection{Details of Post-Processing}
\label{app: post}
\textbf{Segmentation:}
After annotation, we segment the raw recordings into shorter clips to support spatial audio generation and analysis tasks. For speech and singing, we use alignment timestamps to extract utterances. For MRSSound, each recording is divided into fixed 10-second segments.

\textbf{Multimodal Synchronization:}
In addition to binaural audio, each clip is synchronized with the following modalities: audio, text, video, and position data. Using the segment timestamps, we align all streams temporally. For static sources, we attach manually recorded 3D coordinates; for dynamic sources, we interpolate Ultra-Wideband (UWB) tracking data over the segment duration. 
This process yields fully synchronized multimodal clips ready for downstream spatial audio tasks.
Furthermore, for segments where participants' faces are visible, we apply anonymization by using a face detection model to overlay digital masks during post-processing.

\textbf{Checking:}
To ensure the reliability of annotations, domain experts perform a random audit on 15\% of the segmented clips across all four modules. The audit process involves the following steps:  
(1) Verifying the temporal synchronization across different modal;  
(2) Confirming that the assigned event labels accurately reflect the audiovisual content present in each clip;  
(3) Reviewing the speech segments to check the correctness of word- and phoneme-level alignments;  
(4) Evaluating singing clips for accurate alignment between lyrics and audio, consistency with musical scores, and correctness of assigned style labels;  
(5) Assessing MRSMusic excerpts to verify that musical properties, such as key, pitch, and note duration.

\subsection{Statistics of MRSAudio}

\subsubsection{Statistics of MRSLife}
\label{app: MRSLife}
\begin{figure}
\centering
\includegraphics[width=1\linewidth]{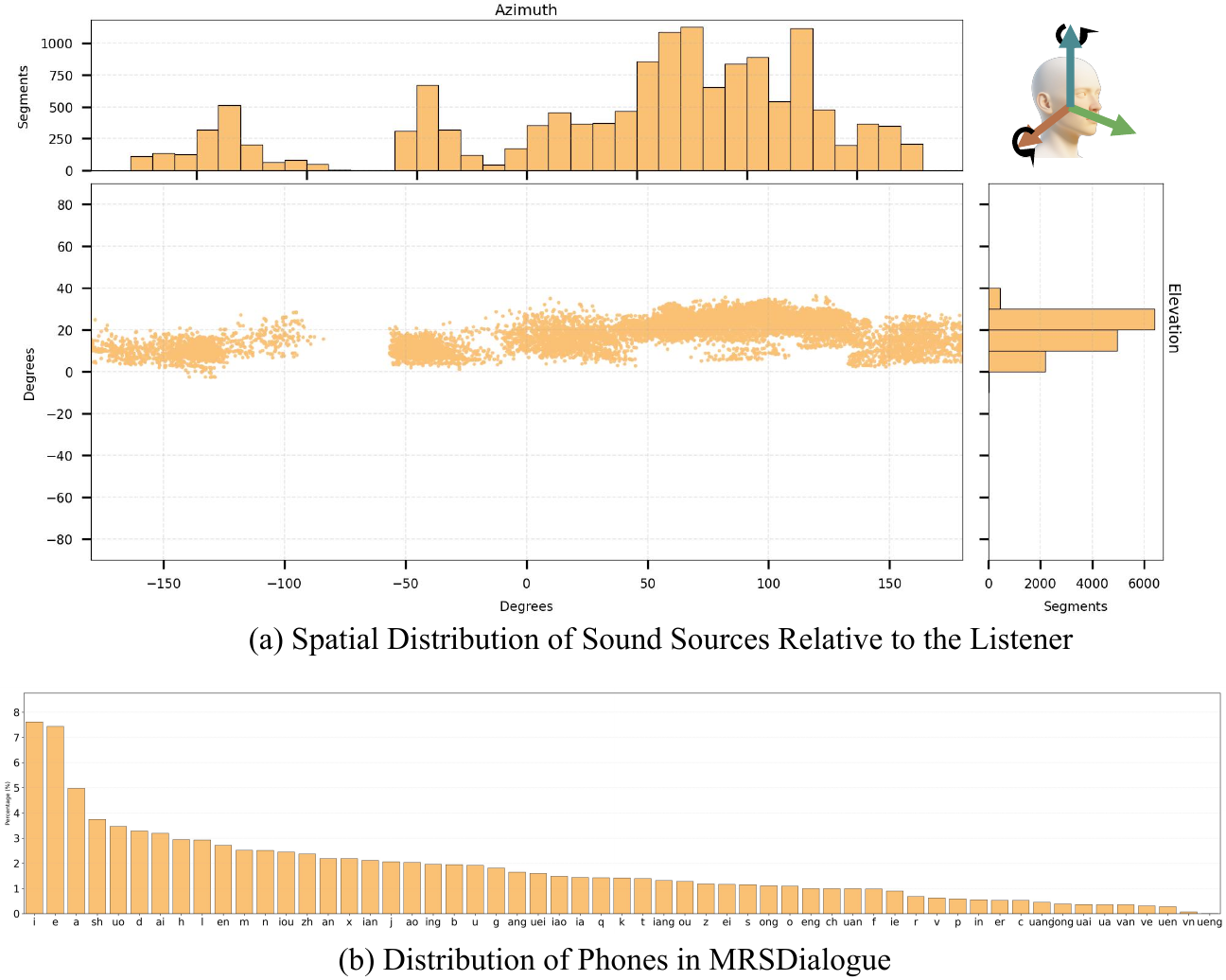}
\caption{
Statistical overview of MRSDialogue. (a) Spatial distribution of sound sources relative to the listener. Red, green, and blue arrows denote the positive x, y, and z axes; azimuth is measured around the z-axis from the x-axis, and elevation is relative to the xy-plane. (b) Distribution of phones in MRSDialogue.
}
\label{fig:stat_dialogue}
\end{figure}
\textbf{MRSDialogue.}  
Figure~\ref{fig:stat_dialogue}(a) presents the 3D spatial distribution of sound sources in MRSDialogue. In this subset, which features frequent human conversations, most sources are located around the ear-level height of the listener. The azimuthal distribution covers nearly all directions surrounding the listener, offering diverse angular data for training spatial localization models with strong generalization.

Figure~\ref{fig:stat_dialogue}(b) shows the phoneme distribution across all speech segments. The most common phoneme is ‘i’, while the least frequent is ‘ueng’. This distribution aligns with real-world phonetic patterns and highlights the dataset’s linguistic richness. Such broad phoneme coverage is beneficial for downstream tasks in speech synthesis and recognition under spatial settings.

\begin{figure}
\centering
\includegraphics[width=1\linewidth]{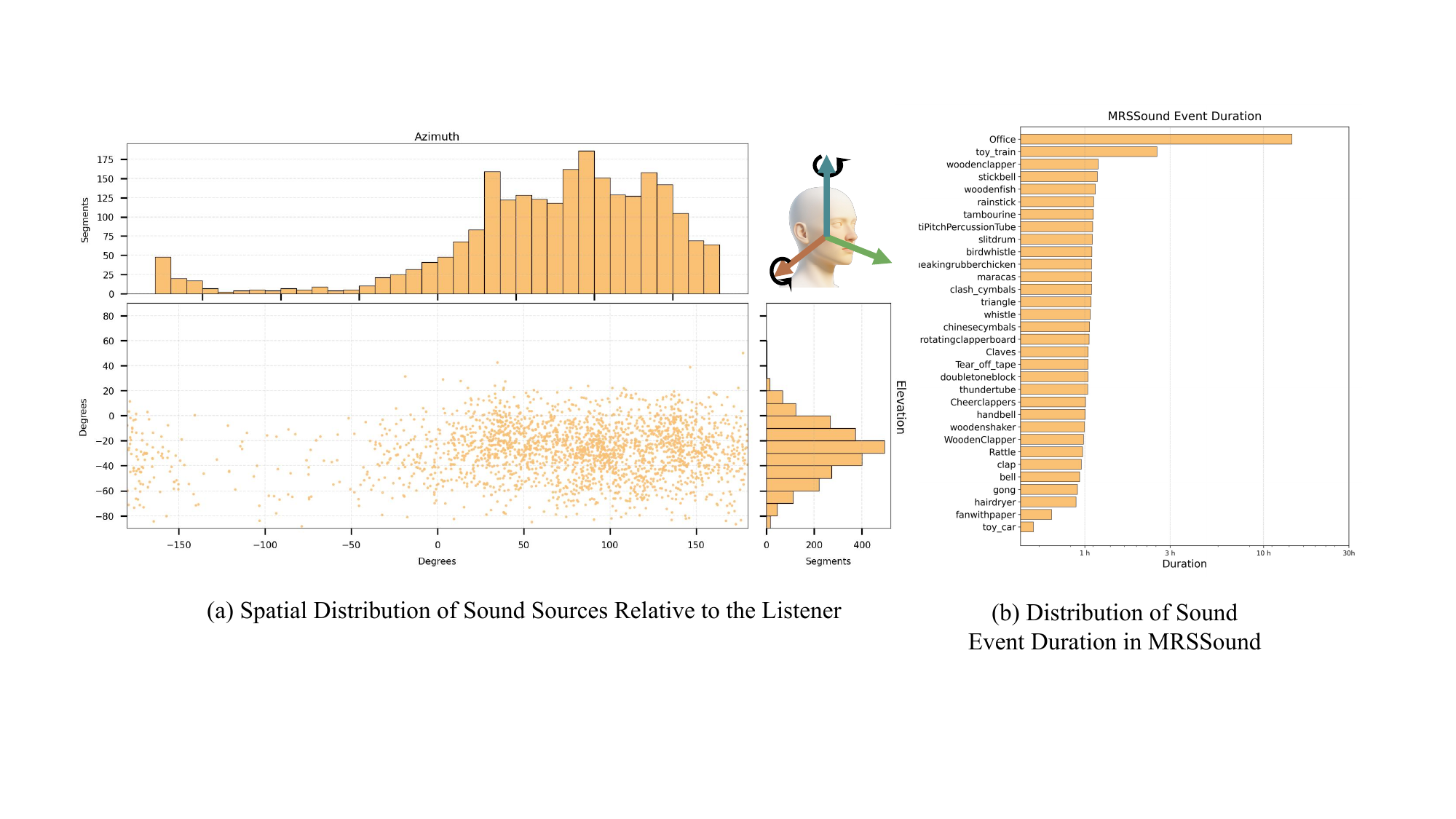}
\caption{
Statistical overview of MRSSound. (a) Spatial distribution of sound sources relative to the listener. Red, green, and blue arrows denote the positive x, y, and z axes; azimuth is measured around the z-axis from the x-axis, and elevation is relative to the xy-plane. (b) Duration distribution of sound event duration in MRSSound.
}
\label{fig:stat_sound}
\end{figure}
\textbf{MRSSound.}  
Figure~\ref{fig:stat_sound}(a) illustrates the spatial distribution of sound sources in MRSSound relative to the listener’s head-centered coordinate system. The listener's facing direction is at 90 degrees, and most sound events are concentrated in the frontal hemisphere (azimuth from 0° to 180°), which is consistent with the first-person video capture setup. Some events also appear in the rear field (–180° to 0°). In elevation, the majority of sound sources are distributed between –90° and 40°, realistically reflecting everyday human perception in standing scenarios, where sound events typically originate below ear level. This broad spatial coverage supports the training of spatial audio models with strong generalization ability.

Figure~\ref{fig:stat_sound}(b) presents the duration distribution of recorded sound events. MRSSound covers a wide variety of everyday scenarios, including cooking in kitchens, working in office environments, and sports-related activities. The diversity of event types and durations makes the subset suitable for training and evaluating models in real-world spatial sound event understanding.

\subsubsection{Statistics of MRSSpeech}
\label{app: MRSSpeech}

\begin{figure}
\centering
\includegraphics[width=\linewidth]{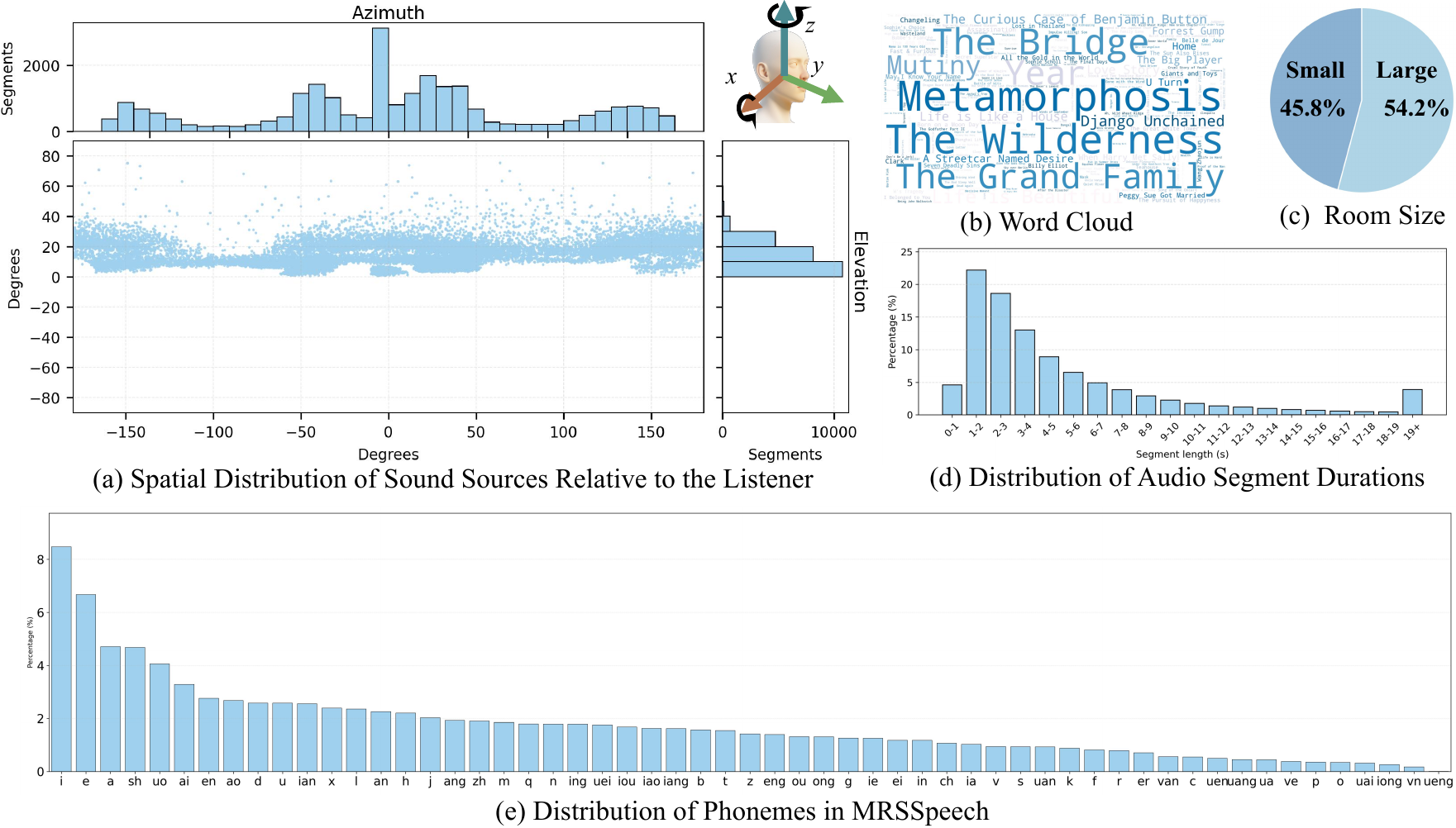}
\caption{
Statistical overview of MRSSpeech. (a) Spatial distribution of sound sources relative to the listener. Red, green, and blue arrows denote the positive x, y, and z axes; azimuth is measured around the z-axis from the x-axis, and elevation is relative to the xy-plane. (b) Word cloud. (c) Proportions of recording spaces by room size. (d) Distribution of audio segment durations. (e) Distribution of phonemes in MRSSpeech.
}
\label{fig:stat_speech}
\end{figure}
Figure~\ref{fig:stat_speech}(a) illustrates the spatial distribution of speech sources with respect to the listener. The azimuth angles span the full 360° around the listener, providing comprehensive coverage of spatial directions. Elevation angles are mostly concentrated between 0° and 60°. Smaller elevations reflect scenarios where both the speaker and listener are either standing or seated, while larger elevations (above 30°) simulate common real-world speech situations such as meetings, where a standing speaker addresses a seated listener. This diverse spatial coverage supports generalization in spatial speech modeling.

Figure~\ref{fig:stat_speech}(b) shows a word cloud representing the diversity of dialogue content. The transcripts are sourced from theatrical scripts, films, and other spoken-only scenarios, capturing a wide range of expressive and stylistic variation. Figure~\ref{fig:stat_speech}(c) presents the distribution of room sizes used for speech recordings. Most multi-speaker interactions take place in medium to large rooms, such as meeting or lecture spaces. We include three distinct environments with varying absorption properties and dimensions to simulate different acoustic conditions.

Figure~\ref{fig:stat_speech}(d) displays the distribution of audio segment durations. Most conversational turns are short, but the dataset also includes extended monologues exceeding 20 seconds, allowing models to capture both brief interactions and long-range motion or speaker dynamics. 

Finally, Figure~\ref{fig:stat_speech}(e) presents the phoneme distribution, which covers all common phonetic units in the dataset’s target language. This phonetic diversity ensures that MRSSpeech provides strong generalizability for phoneme-aware models in speech synthesis and recognition.

\subsubsection{Statistics of MRSSing}
\label{app: MRSSing}
\begin{figure}[ht]
\centering
\includegraphics[width=\linewidth]{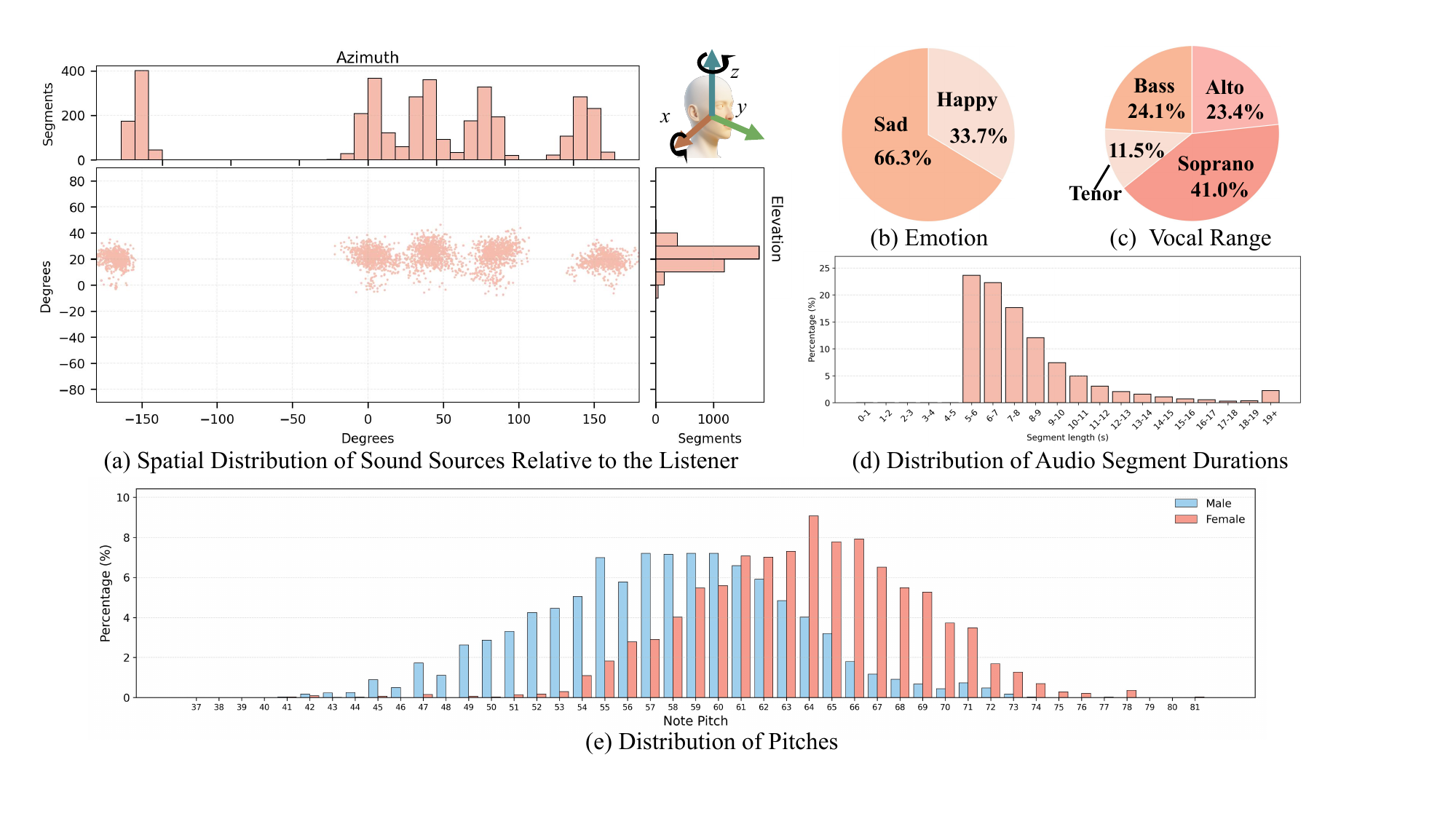}
\caption{
Statistical overview of MRSSing. 
(a) Spatial distribution of sound sources relative to the listener. Red, green, and blue arrows indicate the positive directions of the x, y, and z axes, respectively. Azimuth is measured around the z-axis from the x-axis, and elevation is relative to the xy-plane.
(b) Emotion distribution. 
(c) Vocal range distribution.
(d) Distribution of segment lengths.
(e) Distribution of note pitches across segments.
}
\label{fig:stat_sing}
\end{figure}

Figure~\ref{fig:stat_sing}(a) shows the 3D spatial distribution of sound sources with respect to the listener. The majority of sources are located in front of the listener, consistent with the setup of solo vocal recordings. However, the coverage also spans surrounding directions in both azimuth and elevation, ensuring spatial variability for training robust spatial audio models.

Figures~\ref{fig:stat_sing} (b) and (c) highlight the diversity in style and content. Emotion annotations in Figure~\ref{fig:stat_sing}(b) include expressive labels such as happy and sad. Figure~\ref{fig:stat_sing}(c) displays the coverage across vocal ranges, including soprano, alto, tenor, and bass, ensuring a wide span of pitch and timbral variation.

Figure~\ref{fig:stat_sing}(d) presents the duration distribution of singing segments, which ranges from approximately 4 to 10 seconds. This aligns with typical input lengths used in training singing voice synthesis models. Figure~\ref{fig:stat_sing}(e) illustrates the distribution of note pitches. The full range of musical notes is well represented, and a clear difference in pitch range is observed between male and female singers, with females generally singing at higher pitches. This confirms that MRSSing captures realistic vocal and musical variability.

In summary, MRSSing provides extensive diversity in spatial positioning, language, emotion, vocal range, segment duration, and pitch. This makes it a strong foundation for research on spatial singing voice synthesis and expressive vocal modeling.

\label{app: MRSMusic}
\begin{figure}[ht]
\centering
\includegraphics[width=\linewidth]{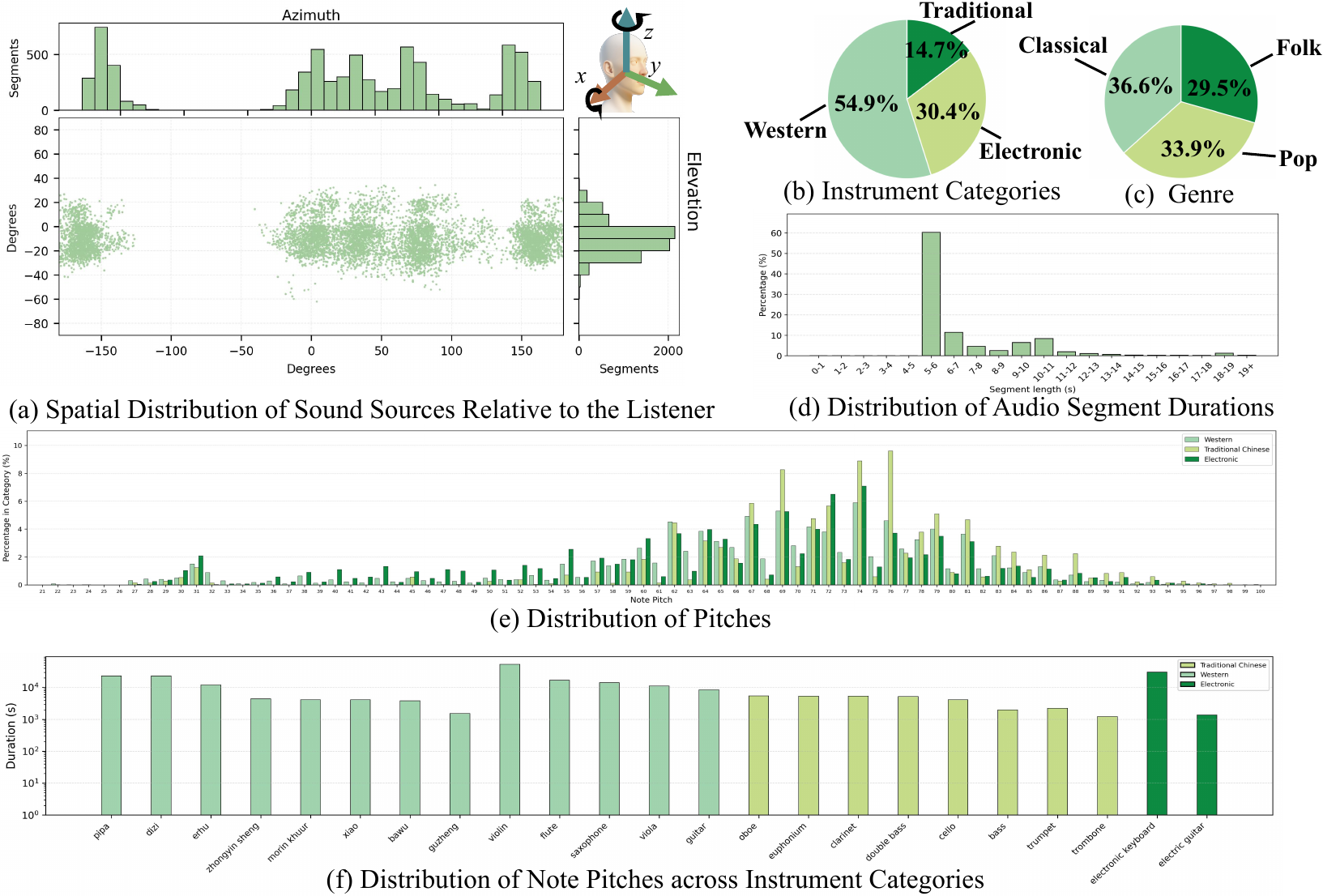}
\caption{
Statistical overview of MRSMusic. (a) Spatial distribution of sound sources relative to the listener. Red, green, and blue arrows denote the positive x, y, and z axes; azimuth is measured around the z-axis from the x-axis, and elevation is relative to the xy-plane. (b) Distribution of instrument categories duration. (c) Distribution of instrument genre duration. (d) Distribution of audio segment durations. (e) Distribution of pitches. (f) Distribution of instrument duration.
}
\label{fig:stat_music}
\end{figure}

\subsubsection{Statistics of MRSMusic}

Figure~\ref{fig:stat_music}(a) illustrates the spatial positions of musical instruments relative to the listener. Most sources are located in front of the listener, consistent with typical music listening scenarios. However, the coverage also includes surrounding directions, contributing to spatial diversity in training and evaluation.

Figures~\ref{fig:stat_music}(b) and (c) highlight the diversity of instrument types and musical genres. The instrument set spans Western, Traditional Chinese, and Electronic categories, while the genre annotations include folk, pop, and classical music. Figure~\ref{fig:stat_music}(f) further details the recording durations across 23 instruments, showing a relatively balanced distribution that facilitates downstream learning for different instrument types.

Regarding generalization capacity, Figure~\ref{fig:stat_music}(d) shows that audio segment durations range from approximately 4 to 11 seconds, matching typical training input lengths. Figure~\ref{fig:stat_music}(e) demonstrates that the dataset covers a full range of musical pitch values, supporting tasks that require robust pitch generalization.

\subsubsection{Statistics of Demographic Representation}
\textbf{Performers.} Across all four subsets, we recruit students from diverse majors at Zhejiang University. The cohort consists of 161 individuals aged between 18 and 25, with a gender ratio of approximately 3:4 (male to female).  

\textbf{Annotation Team.} We employ 25 professional annotators (aged 20–25, gender ratio of about 3:2, male to female) to perform fine-grained annotations of transcripts, phoneme boundaries, music scores, and movement descriptors. All annotators were trained student researchers with relevant domain expertise.  

\section{Details of Experiments}
\label{app: exp}

\subsection{Subjective Evaluation Metrics}
\label{app: sub}
\begin{figure}[ht]
\centering
\includegraphics[width=\linewidth]{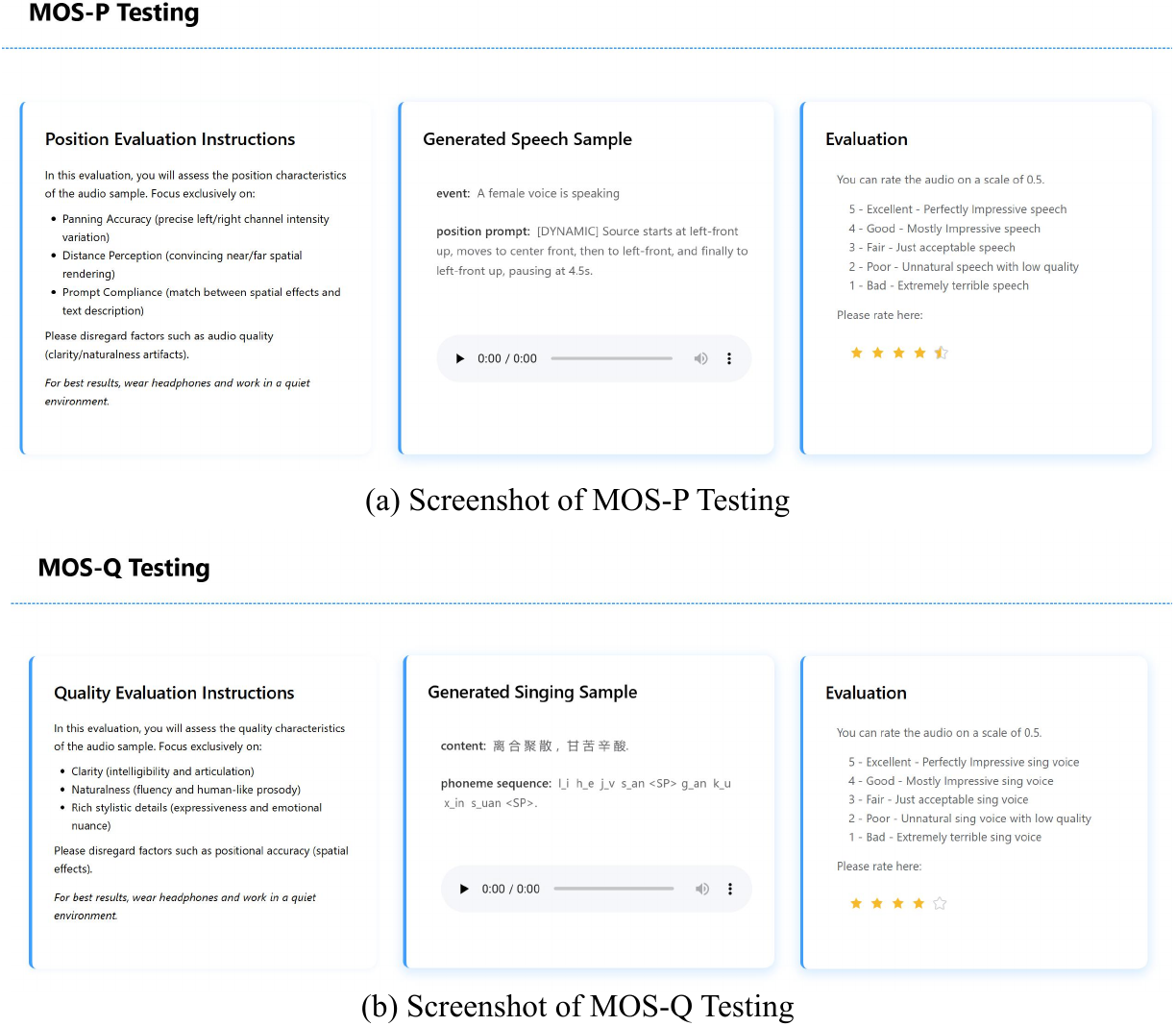}
\caption{
Instructions for audio evaluations. (a) Screenshot of MOS-P Testing. (b) Screenshot of MOS-Q Testing
}
\label{fig:audio_evaluations}
\end{figure}

We conduct subjective evaluation of the generation tasks using Mean Opinion Score (MOS). For each task, we randomly sample 40 utterances from the test set. Each utterance is paired with a corresponding source-position prompt and is evaluated by at least five expert listeners.
MOS-Q, Listeners rate each sample on a five-point Likert scale ranging from 1 (bad) to 5 (excellent).
For audio quality, we use MOS-Q, where listeners wear headphones and assess the clarity and naturalness of the generated audio.
For spatial perception, we use MOS-P, where listeners evaluate the realism of spatial cues and whether the perceived direction and distance of the sound source match the prompt description.
All participants are fairly compensated for their time at a rate of \$15 per hour, resulting in a total cost of approximately \$2000. Participants are informed that their evaluations will be used exclusively for academic research purposes.
Instructions for audio evaluations are shown in Figure
~\ref{fig:audio_evaluations}.

\subsection{Objective Evaluation Metrics}
\label{sec:metrics}

To comprehensively assess spatial audio generation and understanding across multiple tasks, we adopt a set of objective metrics that evaluate signal fidelity, spatial consistency, intelligibility, and speaker or pitch accuracy. We randomly sample 400 data points as the test set.

\paragraph{Audio Spatialization.}  
We measure waveform similarity using Wave L2, the mean squared error (MSE) between the generated and reference binaural waveforms. Amplitude L2 and Phase L2 are computed after applying Short-Time Fourier Transform (STFT), reflecting errors in the magnitude and phase components respectively. MRSTFT loss\footnote{\url{https://github.com/csteinmetz1/auraloss}} is also used, combining spectral convergence, log-magnitude, and linear-magnitude terms for better spectral alignment.
In addition, we use the perceptual evaluation metric PESQ\footnote{\url{https://github.com/aliutkus/speechmetrics}} to assess audio quality for speech-related tasks. Since PESQ is designed for speech, it is omitted for MRSLife and MRSMusic. For all metrics except PESQ, lower values indicate better performance.

To evaluate spatial consistency, we use Spatial-AST\citep{zheng2024bat} to extract angular and distance embeddings from the binaural audio. Since Spatial-AST predicts positions only for static sources, we compute the cosine similarity between the predicted and ground truth embeddings within 1-second segments and average the results to assess overall spatial fidelity.

\paragraph{Spatial Text to Speech.}  
We evaluate speech intelligibility using Character Error Rate (CER), which measures the proportion of character-level differences between ASR transcriptions and reference texts. Transcriptions are generated using the Paraformer-zh model~\citep{gao2023funasr}. To assess speaker consistency, we compute Speaker Identity Matching (SIM) as the cosine similarity between speaker embeddings extracted using a WavLM-based speaker verification model.\footnote{\url{https://huggingface.co/pyannote/speaker-diarization}}

\paragraph{Spatial Singing Voice Synthesis.}  
We use Mel Cepstral Distortion (MCD) to assess the spectral similarity between the generated and reference vocals. It is defined as:

\begin{equation}
\text{MCD} = \frac{10}{\ln 10} \sqrt{2 \sum_{d=1}^{D} (m_t(d) - \hat{m}_t(d))^2},
\end{equation}

where $m_t(d)$ and $\hat{m}_t(d)$ represent the $d$-th Mel-frequency cepstral coefficient (MFCC) at frame $t$ for the ground truth and synthesized signals, respectively, and $D$ is the number of MFCC dimensions.

Additionally, we adopt F0 Frame Error (FFE) to evaluate pitch accuracy by comparing extracted F0 contours between the synthesized and ground truth audio.

\paragraph{Spatial Music Generation.}  
We use Fréchet Audio Distance (FAD)~\citep{kilgour2018fr} to assess perceptual similarity between the feature distributions of generated and reference audio. In addition, F0 Frame Error (FFE) is used to evaluate pitch accuracy by comparing the extracted F0 contours against the reference musical scores.

\paragraph{Sound Event Localization and Detection (SELD).}  
Following STARSS23~\citep{shimada2023starss23}, we adopt four joint detection and localization metrics:
F$_{20^\circ}$: location-aware F-score; a prediction is correct if the event class matches and angular error is below $20^\circ$.
ER$_{20^\circ}$: error rate computed as the sum of insertions, deletions, and substitutions over reference events.
LE$_{CD}$: class-aware localization error, the mean angular difference between predicted and reference directions.
LR$_{CD}$: class-aware localization recall, the percentage of correctly localized events among all instances of each class.
We compute all metrics in 1-second non-overlapping segments using macro-averaging across all event classes. Higher values of F$_{20^\circ}$ and LR$_{CD}$, and lower values of ER$_{20^\circ}$ and LE$_{CD}$ indicate better performance.

\subsection{Audio Spatialization}
\label{app:bas}
We build upon BinauralGrad’s two-stage diffusion-based framework to convert monaural audio into spatial audio. However, in our subsequent generation experiments, the synthesized monaural audio typically corresponds to a centrally positioned source between the ears, effectively serving as the first stage of BinauralGrad. Therefore, for the spatialization experiments presented here, we use only the second stage of BinauralGrad to validate spatial audio generation.
Specifically, we first convert binaural recordings to monaural input by averaging the two channels. Next, we apply a DSP-based method to produce a coarse spatial approximation of the binaural signal. This monaural input and its simulated binaural counterpart are then used as input to the BinauralGrad model, which is conditioned on the object's motion trajectory to generate spatialized binaural audio.
We list the architecture and hyperparameters of BinauralGrad in Table \ref{tab: bingrad_arch}.

\begin{table}[ht]
\centering
\caption{Hyper-parameters of BinauralGrad modules.}
\begin{tabular}{l|c|c}
\toprule
\multicolumn{2}{c|}{\bfseries{Hyperparameter}} & \bfseries{BinauralGrad} \\
\midrule
\multirow{4}{*}{\shortstack{Binaural\\Encoder}}
~ & Wave Encoder Layers & 2 \\
~ & Position Encoder Layers & 2 \\
~ & Encoder Conv1D Kernel & 3 \\
~ & Encoder Dropout & 0.4 \\
\midrule
\multirow{5}{*}{\shortstack{Mel Predictor}} & Residual Blocks & 3 \\
~ & Bidirectional Layers & 3 \\
~ & Hidden\_size & 128 \\
~ & Training Steps & 200 \\
~ & Sampling Steps & 6 \\
\bottomrule
\end{tabular}
\label{tab: bingrad_arch}
\end{table}
\subsection{Spatial Text to Speech}
\label{app:btts}
We fine-tune two pre-trained models, CosyVoice and F5-TTS, using monaural audio obtained by averaging the binaural recordings from the MRSSpeech subset. This allows the models to learn the generation characteristics specific to our dataset. After monaural generation, we apply a pre-trained spatialization model trained on MRSSpeech to convert the output into spatialized binaural audio. For the ISDrama baseline, we adopt the model variant proposed in the original paper that conditions generation on predefined spatial paths to produce binaural spatial speech directly.

\begin{table}[ht]
\centering
\caption{Hyper-parameters of Rmssinger modules.}
\begin{tabular}{l|c|c}
\toprule
\multicolumn{2}{c|}{\bfseries{Hyperparameter}} & \bfseries{Rmssinger} \\
\midrule
\multirow{7}{*}{\shortstack{Phoneme\\Encoder}} & Phoneme Embedding & 256 \\
~ & Encoder Layers & 4 \\
~ & Encoder Hidden & 256 \\
~ & Encoder Conv1D Kernel & 9 \\
~ & Encoder Conv1D Filter Size & 1024 \\
~ & Encoder Attention Heads & 2 \\
~ & Encoder Dropout & 0.1 \\
\midrule
\multirow{3}{*}{\shortstack{Note\\Encoder}} & Pitches Embedding & 256 \\
~ & Type Embedding & 256 \\
~ & Duration Hidden & 256 \\
\midrule
\multirow{5}{*}{\shortstack{Pitch\\Predictor}} & Conv Layers & 12 \\
~ & Kernel Size & 3 \\
~ & Residual Channel & 192 \\
~ & Hidden Channel & 256 \\
~ & Training Steps & 100 \\
\midrule
\multirow{5}{*}{\shortstack{Mel Predictor}} & Conv Layers & 20 \\
~ & Kernel Size & 3 \\
~ & Residual Channel & 256 \\
~ & Hidden Channel & 256 \\
~ & Training Steps & 100 \\
\bottomrule
\end{tabular}

\label{tab: vecarch}
\end{table}

\subsection{Spatial Singing Voice Synthesis}
\label{app:bsvs}
For the ISDrama baseline, we adopt the spatial-path-conditioned generation framework and extend it by incorporating Rmssinger’s note encoder, allowing explicit pitch control through musical score input.
We use the single stage variation of Rmssinger, a standard singing voice synthesis model, and train it with monaural targets derived from the averaged binaural recordings in the MRSSing subset. This enables the model to generate pitch-accurate audio that reflects the acoustic characteristics. The synthesized monaural audio is then spatialized using the spatialization model trained on MRSSing. 
We list the architecture and hyperparameters of Rmssinger in Table \ref{tab: vecarch}.

\subsection{Spatial Music Generation}
\label{app:smg}
For spatial music generation, in the ISDrama setting, we follow the path-conditioned generation pipeline but remove the phoneme encoder, using only the note encoder to process symbolic scores as the sole control condition for generating spatialized music.
We also adopt Make-An-Audio 2 as our base model and train it using monaural audio derived from averaged MRSMusic binaural recordings. We use the pretrained spectrogram autoencoder. Instrument category and symbolic score information are concatenated into the text prompt to guide the music generation process. The generated audio is subsequently spatialized using the MRSMusic-trained spatialization model. We list the hyper-parameters of Make-An-Audio 2 in Table~\ref{tab:hyper}. 

\begin{table}[ht]
\centering
\caption{Hyperparameters of Make-An-Audio 2.}
\begin{tabular}{l|c|c}
\toprule
\multicolumn{2}{c|}{Hyperparameter}   & Make-An-Audio 2 \\ 
\midrule
\multirow{5}{*}{Autoencoders}
&Input/Output Channels                    & 80      \\
&Hidden Channels                          &   20   \\ 
&Residual Blocks                         &   2   \\   
&Spectrogram Shape                &  (80, 624) \\    
&Channel Multipiler                &   $[1, 2, 4]$ \\   
\midrule
\multirow{6}{*}{Transformer Backbone}
& Input shape       &  (20, T)   \\         
& Condition\_embedding Size              &  1024 \\ 
& Feed-forward Hidden\_size             &  576 \\
& Num of Transformer Heads          &  8 \\
& Transformer Blocks                 &  8\\
& Training Steps                 &  1000\\
& Sampling Steps                 &  100\\
\midrule
\multirow{3}{*}{CLAP Text Encoder}
&Transformer Embed Channels        &  768   \\         
&Output Project Channels   &  1024 \\    
&Token Length   &  77 \\         
\bottomrule
\end{tabular}

\label{tab:hyper}
\end{table}

\begin{table}[ht]
\centering
\caption{Hyperparameters of SELD.}
\begin{tabular}{l|c|c}
\toprule
\multicolumn{2}{c|}{Hyperparameter}   & SELD \\ 
\midrule
\multirow{4}{*}{Input}
&Binaural Channels                    & 3      \\
&FOA Channels                          &   7   \\ 
&Frequency Bins                         &   128   \\   
&Frames                &  120 \\    
\midrule
\multirow{2}{*}{Audio Encoder(CNN)}
& Hidden\_size             &  64 \\
& Conv Blocks                 &  3\\
\midrule
\multirow{3}{*}{Audio Encoder(Vit)}
& Hidden\_size             &  128 \\
& Conv Blocks                 &  1\\
& Transformer Blocks                 &  2\\   
& Num of Transformer Heads             &  4 \\
\bottomrule
\end{tabular}

\label{tab:hyper_seld}
\end{table}

\subsection{Sound Event Localization and Detection}
\label{app:seld}
We follow the STARSS23 framework for FOA-based sound event localization and detection, training on the event segments from the MRSSound subset under both audio-only and audio-visual conditions. To enable binaural input, we modify the model architecture to use three input channels and extract interaural phase difference features. This allows us to adapt the SELD model to binaural audio. Additionally, we experiment with replacing convolutional layers with Transformer encoders to explore the effect of different architectures on sound event localization and detection performance.
We list the hyper-parameters of SELD in Table~\ref{tab:hyper_seld}. 

\clearpage

%% file: neurips_2025.bbl
\begin{thebibliography}{44}
\providecommand{\natexlab}[1]{#1}
\providecommand{\url}[1]{\texttt{#1}}
\expandafter\ifx\csname urlstyle\endcsname\relax
  \providecommand{\doi}[1]{doi: #1}\else
  \providecommand{\doi}{doi: \begingroup \urlstyle{rm}\Url}\fi

\bibitem[Achiam et~al.(2023)Achiam, Adler, Agarwal, Ahmad, Akkaya, Aleman, Almeida, Altenschmidt, Altman, Anadkat, et~al.]{achiam2023gpt}
Achiam, J., Adler, S., Agarwal, S., Ahmad, L., Akkaya, I., Aleman, F.~L., Almeida, D., Altenschmidt, J., Altman, S., Anadkat, S., et~al.
\newblock Gpt-4 technical report.
\newblock \emph{arXiv preprint arXiv:2303.08774}, 2023.

\bibitem[Adavanne et~al.(2019)Adavanne, Politis, and Virtanen]{Adavanne2019_DCASE}
Adavanne, S., Politis, A., and Virtanen, T.
\newblock A multi-room reverberant dataset for sound event localization and detection.
\newblock \emph{Proc. DCASE2019}, 2019.

\bibitem[Agostinelli et~al.(2023)Agostinelli, Denk, Borsos, Engel, Verzetti, Caillon, Huang, Jansen, Roberts, Tagliasacchi, et~al.]{agostinelli2023musiclm}
Agostinelli, A., Denk, T.~I., Borsos, Z., Engel, J., Verzetti, M., Caillon, A., Huang, Q., Jansen, A., Roberts, A., Tagliasacchi, M., et~al.
\newblock Musiclm: Generating music from text.
\newblock \emph{arXiv preprint arXiv:2301.11325}, 2023.

\bibitem[Aiello \& Rogerson(2003)Aiello and Rogerson]{aiello2003ultra}
Aiello, G.~R. and Rogerson, G.~D.
\newblock Ultra-wideband wireless systems.
\newblock \emph{IEEE microwave magazine}, 4\penalty0 (2):\penalty0 36--47, 2003.

\bibitem[Baevski et~al.(2020)Baevski, Zhou, Mohamed, and Auli]{baevski2020wav2vec}
Baevski, A., Zhou, Y., Mohamed, A., and Auli, M.
\newblock wav2vec 2.0: A framework for self-supervised learning of speech representations.
\newblock \emph{Advances in neural information processing systems}, 33:\penalty0 12449--12460, 2020.

\bibitem[Bain et~al.(2023)Bain, Huh, Han, and Zisserman]{bain2023whisperx}
Bain, M., Huh, J., Han, T., and Zisserman, A.
\newblock Whisperx: Time-accurate speech transcription of long-form audio.
\newblock \emph{arXiv preprint arXiv:2303.00747}, 2023.

\bibitem[Bittner et~al.(2022)Bittner, Bosch, Rubinstein, Meseguer-Brocal, and Ewert]{2022_BittnerBRME_LightweightNoteTranscription_ICASSP}
Bittner, R.~M., Bosch, J.~J., Rubinstein, D., Meseguer-Brocal, G., and Ewert, S.
\newblock A lightweight instrument-agnostic model for polyphonic note transcription and multipitch estimation.
\newblock In \emph{Proceedings of the IEEE International Conference on Acoustics, Speech, and Signal Processing (ICASSP)}, Singapore, 2022.

\bibitem[Boersma(2001)]{boersma2001praat}
Boersma, P.
\newblock Praat, a system for doing phonetics by computer.
\newblock \emph{Glot. Int.}, 5\penalty0 (9):\penalty0 341--345, 2001.

\bibitem[Chen et~al.(2020)Chen, Xie, Vedaldi, and Zisserman]{chen2020vggsound}
Chen, H., Xie, W., Vedaldi, A., and Zisserman, A.
\newblock Vggsound: A large-scale audio-visual dataset.
\newblock In \emph{ICASSP 2020-2020 IEEE International Conference on Acoustics, Speech and Signal Processing (ICASSP)}, pp.\  721--725. IEEE, 2020.

\bibitem[Chen et~al.(2024)Chen, Niu, Ma, Deng, Wang, Zhao, Yu, and Chen]{chen2024f5}
Chen, Y., Niu, Z., Ma, Z., Deng, K., Wang, C., Zhao, J., Yu, K., and Chen, X.
\newblock F5-tts: A fairytaler that fakes fluent and faithful speech with flow matching.
\newblock \emph{arXiv preprint arXiv:2410.06885}, 2024.

\bibitem[Chu et~al.(2023)Chu, Xu, Zhou, Yang, Zhang, Yan, Zhou, and Zhou]{Qwen-Audio}
Chu, Y., Xu, J., Zhou, X., Yang, Q., Zhang, S., Yan, Z., Zhou, C., and Zhou, J.
\newblock Qwen-audio: Advancing universal audio understanding via unified large-scale audio-language models.
\newblock \emph{arXiv preprint arXiv:2311.07919}, 2023.

\bibitem[Cohen \& Knudsen(1999)Cohen and Knudsen]{cohen1999maps}
Cohen, Y.~E. and Knudsen, E.~I.
\newblock Maps versus clusters: different representations of auditory space in the midbrain and forebrain.
\newblock \emph{Trends in neurosciences}, 22\penalty0 (3):\penalty0 128--135, 1999.

\bibitem[Copet et~al.(2023)Copet, Kreuk, Gat, Remez, Kant, Synnaeve, Adi, and Défossez]{copet2023simple}
Copet, J., Kreuk, F., Gat, I., Remez, T., Kant, D., Synnaeve, G., Adi, Y., and Défossez, A.
\newblock Simple and controllable music generation.
\newblock \emph{arXiv preprint arXiv:2306.05284}, 2023.

\bibitem[Du et~al.(2024)Du, Chen, Zhang, Hu, Lu, Yang, Hu, Zheng, Gu, Ma, et~al.]{du2024cosyvoice}
Du, Z., Chen, Q., Zhang, S., Hu, K., Lu, H., Yang, Y., Hu, H., Zheng, S., Gu, Y., Ma, Z., et~al.
\newblock Cosyvoice: A scalable multilingual zero-shot text-to-speech synthesizer based on supervised semantic tokens.
\newblock \emph{arXiv preprint arXiv:2407.05407}, 2024.

\bibitem[Gao \& Grauman(2019)Gao and Grauman]{Gao20192.5D}
Gao, R. and Grauman, K.
\newblock 2.5d visual sound.
\newblock \emph{arXiv preprint arXiv:1812.04204}, 2019.

\bibitem[Gao et~al.(2023)Gao, Li, Wang, Luo, Shi, Chen, Li, Zuo, Du, Xiao, et~al.]{gao2023funasr}
Gao, Z., Li, Z., Wang, J., Luo, H., Shi, X., Chen, M., Li, Y., Zuo, L., Du, Z., Xiao, Z., et~al.
\newblock Funasr: A fundamental end-to-end speech recognition toolkit.
\newblock \emph{arXiv preprint arXiv:2305.11013}, 2023.

\bibitem[Gemmeke et~al.(2017)Gemmeke, Ellis, Freedman, Jansen, Lawrence, Moore, Plakal, and Ritter]{jort_audioset_2017}
Gemmeke, J.~F., Ellis, D. P.~W., Freedman, D., Jansen, A., Lawrence, W., Moore, R.~C., Plakal, M., and Ritter, M.
\newblock Audio set: An ontology and human-labeled dataset for audio events.
\newblock In \emph{Proc. IEEE ICASSP 2017}, New Orleans, LA, 2017.

\bibitem[Grothe et~al.(2010)Grothe, Pecka, and McAlpine]{grothe2010mechanisms}
Grothe, B., Pecka, M., and McAlpine, D.
\newblock Mechanisms of sound localization in mammals.
\newblock \emph{Physiological reviews}, 90\penalty0 (3):\penalty0 983--1012, 2010.

\bibitem[He et~al.(2023)He, Liu, Ye, Huang, Cui, Liu, and Zhao]{he2023rmssinger}
He, J., Liu, J., Ye, Z., Huang, R., Cui, C., Liu, H., and Zhao, Z.
\newblock Rmssinger: Realistic-music-score based singing voice synthesis.
\newblock \emph{arXiv preprint arXiv:2305.10686}, 2023.

\bibitem[Huang et~al.(2023{\natexlab{a}})Huang, Ren, Huang, Yang, Ye, Zhang, Liu, Yin, Ma, and Zhao]{huang2023makeanaudio}
Huang, J., Ren, Y., Huang, R., Yang, D., Ye, Z., Zhang, C., Liu, J., Yin, X., Ma, Z., and Zhao, Z.
\newblock Make-an-audio 2: Temporal-enhanced text-to-audio generation, 2023{\natexlab{a}}.

\bibitem[Huang et~al.(2023{\natexlab{b}})Huang, Li, Yang, Shi, Chang, et~al.]{huang2023audiogpt}
Huang, R., Li, M., Yang, D., Shi, J., Chang, X., et~al.
\newblock {AudioGPT}: Understanding and generating speech, music, sound, and talking head.
\newblock \emph{arXiv preprint arXiv:2304.12995}, 2023{\natexlab{b}}.

\bibitem[Huiyu et~al.(2025)Huiyu, Shuaifan, Zhibo, Zhongjie, and Tao]{huiyu2025psa}
Huiyu, X., Shuaifan, J., Zhibo, W., Zhongjie, B., and Tao, W.
\newblock Psa-nerf: Personalized spatial attention neural rendering for audio-driven talking portraits generation.
\newblock \emph{Chinese Journal of Electronics}, 2025.

\bibitem[Kilgour et~al.(2018)Kilgour, Zuluaga, Roblek, and Sharifi]{kilgour2018fr}
Kilgour, K., Zuluaga, M., Roblek, D., and Sharifi, M.
\newblock Fr$\backslash$'echet audio distance: A metric for evaluating music enhancement algorithms.
\newblock \emph{arXiv preprint arXiv:1812.08466}, 2018.

\bibitem[Kim et~al.(2025)Kim, Yun, and Kim]{kimvisage}
Kim, J., Yun, H., and Kim, G.
\newblock Visage: Video-to-spatial audio generation.
\newblock In \emph{ICLR}, 2025.

\bibitem[Kreuk et~al.(2022)Kreuk, Synnaeve, Polyak, Singer, D{\'e}fossez, Copet, Parikh, Taigman, and Adi]{kreuk2022audiogen}
Kreuk, F., Synnaeve, G., Polyak, A., Singer, U., D{\'e}fossez, A., Copet, J., Parikh, D., Taigman, Y., and Adi, Y.
\newblock Audiogen: Textually guided audio generation.
\newblock \emph{arXiv preprint arXiv:2209.15352}, 2022.

\bibitem[Leng et~al.(2022)Leng, Chen, Guo, Liu, Chen, Tan, Mandic, He, Li, Qin, et~al.]{leng2022binauralgrad}
Leng, Y., Chen, Z., Guo, J., Liu, H., Chen, J., Tan, X., Mandic, D., He, L., Li, X., Qin, T., et~al.
\newblock Binauralgrad: A two-stage conditional diffusion probabilistic model for binaural audio synthesis.
\newblock \emph{Advances in Neural Information Processing Systems}, 35:\penalty0 23689--23700, 2022.

\bibitem[Li et~al.(2024)Li, Zhang, Wang, Hong, Huang, and Zhao]{li2024robust}
Li, R., Zhang, Y., Wang, Y., Hong, Z., Huang, R., and Zhao, Z.
\newblock Robust singing voice transcription serves synthesis, 2024.

\bibitem[Liu et~al.(2025)Liu, Luo, Jiang, Luo, Sun, Wan, Huang, Chen, Wang, Li, Zhang, Yan, Zhao, and Xue]{liu2025omniaudiogeneratingspatialaudio}
Liu, H., Luo, T., Jiang, Q., Luo, K., Sun, P., Wan, J., Huang, R., Chen, Q., Wang, W., Li, X., Zhang, S., Yan, Z., Zhao, Z., and Xue, W.
\newblock Omniaudio: Generating spatial audio from 360-degree video, 2025.
\newblock URL \url{https://arxiv.org/abs/2504.14906}.

\bibitem[Liu et~al.(2022)Liu, Rosen, et~al.]{liu2022autoslicer}
Liu, Z., Rosen, E., et~al.
\newblock Autoslicer: Scalable automated data slicing for ml model analysis.
\newblock \emph{arXiv preprint arXiv:2212.09032}, 2022.

\bibitem[McAuliffe et~al.(2017)McAuliffe, Socolof, Mihuc, Wagner, and Sonderegger]{mcauliffe2017montreal}
McAuliffe, M., Socolof, M., Mihuc, S., Wagner, M., and Sonderegger, M.
\newblock Montreal forced aligner: Trainable text-speech alignment using kaldi.
\newblock In \emph{Interspeech}, volume 2017, pp.\  498--502, 2017.

\bibitem[Pedro~Morgado \& Wang(2018)Pedro~Morgado and Wang]{morgadoNIPS18}
Pedro~Morgado, Nuno~Vasconcelos, T.~L. and Wang, O.
\newblock Self-supervised generation of spatial audio for 360$\deg$ video.
\newblock In \emph{Neural Information Processing Systems (NIPS)}, 2018.

\bibitem[Sarabia et~al.(2023)Sarabia, Menyaylenko, Toso, Seto, Aldeneh, Pirhosseinloo, Zappella, Theobald, Apostoloff, and Sheaffer]{sarabia2023spatial}
Sarabia, M., Menyaylenko, E., Toso, A., Seto, S., Aldeneh, Z., Pirhosseinloo, S., Zappella, L., Theobald, B.-J., Apostoloff, N., and Sheaffer, J.
\newblock Spatial librispeech: An augmented dataset for spatial audio learning.
\newblock \emph{arXiv preprint arXiv:2308.09514}, 2023.

\bibitem[Shimada et~al.(2023)Shimada, Politis, Sudarsanam, Krause, Uchida, Adavanne, Hakala, Koyama, Takahashi, Takahashi, et~al.]{shimada2023starss23}
Shimada, K., Politis, A., Sudarsanam, P., Krause, D.~A., Uchida, K., Adavanne, S., Hakala, A., Koyama, Y., Takahashi, N., Takahashi, S., et~al.
\newblock Starss23: An audio-visual dataset of spatial recordings of real scenes with spatiotemporal annotations of sound events.
\newblock \emph{Advances in neural information processing systems}, 36:\penalty0 72931--72957, 2023.

\bibitem[Sun et~al.(2024)Sun, Cheng, Li, Ye, Liu, Zhang, Xue, and Guo]{sun2024both}
Sun, P., Cheng, S., Li, X., Ye, Z., Liu, H., Zhang, H., Xue, W., and Guo, Y.
\newblock Both ears wide open: Towards language-driven spatial audio generation.
\newblock \emph{arXiv preprint arXiv:2410.10676}, 2024.

\bibitem[Tang et~al.(2024)Tang, Yu, Sun, Chen, Tan, et~al.]{tang2023salmonn}
Tang, C., Yu, W., Sun, G., Chen, X., Tan, T., et~al.
\newblock {SALMONN}: Towards generic hearing abilities for large language models.
\newblock \emph{Proc. ICLR}, 2024.

\bibitem[Wang et~al.(2022)Wang, Chai, Wu, Nian, Niu, Zheng, Wang, Sun, Fang, Pan, et~al.]{wang2022nerc}
Wang, Q., Chai, L., Wu, H., Nian, Z., Niu, S., Zheng, S., Wang, Y., Sun, L., Fang, Y., Pan, J., et~al.
\newblock The nerc-slip system for sound event localization and detection of dcase2022 challenge.
\newblock \emph{DCASE2022 Challenge, Tech. Rep.}, 2022.

\bibitem[Wang et~al.(2024)Wang, Guo, Huang, Huang, Wang, You, Li, and Zhao]{wang2024frieren}
Wang, Y., Guo, W., Huang, R., Huang, J., Wang, Z., You, F., Li, R., and Zhao, Z.
\newblock Frieren: Efficient video-to-audio generation with rectified flow matching.
\newblock \emph{arXiv e-prints}, pp.\  arXiv--2406, 2024.

\bibitem[Wei et~al.(2023)Wei, Cao, Dan, and Chen]{wei2023rmvpe}
Wei, H., Cao, X., Dan, T., and Chen, Y.
\newblock Rmvpe: A robust model for vocal pitch estimation in polyphonic music.
\newblock \emph{arXiv preprint arXiv:2306.15412}, 2023.

\bibitem[Xie(2020)]{xie2020spatial}
Xie, B.
\newblock Spatial sound-history, principle, progress and challenge.
\newblock \emph{Chinese Journal of Electronics}, 29\penalty0 (3):\penalty0 397--416, 2020.

\bibitem[Yang et~al.(2024)Yang, Quan, Wang, Wang, Yang, Fang, Shao, Bu, Xu, and Li]{yang2024realman}
Yang, B., Quan, C., Wang, Y., Wang, P., Yang, Y., Fang, Y., Shao, N., Bu, H., Xu, X., and Li, X.
\newblock Realman: A real-recorded and annotated microphone array dataset for dynamic speech enhancement and localization.
\newblock \emph{Advances in Neural Information Processing Systems}, 37:\penalty0 105997--106019, 2024.

\bibitem[Yang et~al.(2023)Yang, Tian, Tan, Huang, Liu, Chang, Shi, Zhao, Bian, Wu, et~al.]{yang2023uniaudio}
Yang, D., Tian, J., Tan, X., Huang, R., Liu, S., Chang, X., Shi, J., Zhao, S., Bian, J., Wu, X., et~al.
\newblock Uniaudio: An audio foundation model toward universal audio generation.
\newblock \emph{arXiv preprint arXiv:2310.00704}, 2023.

\bibitem[Yost(1998)]{yost1998spatial}
Yost, W.~A.
\newblock Spatial hearing: the psychophysics of human sound localization.
\newblock \emph{Ear and Hearing}, 19\penalty0 (2):\penalty0 167, 1998.

\bibitem[Zhang et~al.(2025)Zhang, Guo, Pan, Zhu, Jin, and Zhao]{zhang2025isdrama}
Zhang, Y., Guo, W., Pan, C., Zhu, Z., Jin, T., and Zhao, Z.
\newblock Isdrama: Immersive spatial drama generation through multimodal prompting.
\newblock \emph{arXiv preprint arXiv:2504.20630}, 2025.

\bibitem[Zheng et~al.(2024)Zheng, Peng, Ma, Chen, Choi, and Harwath]{zheng2024bat}
Zheng, Z., Peng, P., Ma, Z., Chen, X., Choi, E., and Harwath, D.
\newblock Bat: Learning to reason about spatial sounds with large language models.
\newblock \emph{arXiv preprint arXiv:2402.01591}, 2024.

\end{thebibliography}
